\documentclass[apj,twocolumn,twocolappendix,numberedappendix]{openjournal} \usepackage[breaklinks,colorlinks,citecolor=magenta,urlcolor=blue,linkcolor=blue,filecolor=blue]{hyperref}

\newcommand{\ha}{H$\alpha$}

\newcommand{\orcid}[1]{\href{https://orcid.org/#1}{\textcolor[HTML]{A6CE39}{\aiOrcid}}}

\def\ltsima{$\buildrel<\over\sim$}
\def\lsim{\lower.5ex\hbox{\ltsima}~}
\def\gtsima{$\buildrel>\over\sim$}
\def\gsim{\lower.5ex\hbox{\gtsima}~}

\def\rhouv{$\rho_{\mathrm{UV}}$}
\newcommand{\muv}{$M_{\mathrm {UV}}$}

\def\msol{M$_{\odot}$}

\def\teff{\ifmmode T_{\rm eff} \else $T_{\mathrm{eff}}$\fi}

\def\lya{Ly$\alpha$} 
\def\ha{H$\alpha$}

\newcommand{\fesc}{$f_{\mathrm{esc}}$}

\def\cm2{cm$^{-2}$}

\def\ewo3{$EW_{\mathrm{[O\textsc{iii}]}}$}

\def\nh{\ifmmode N_{\mathrm{HI}}\else $N_{\mathrm{HI}}$\fi}

\def\vexp{\ifmmode v_{\mathrm exp} \else v$_{\mathrm exp}$\fi}
\def\taua{\ifmmode \tau_{a}\else $\tau_{a}$\fi}

\def\xiion{$\xi_{\mathrm{ion}}$}
\def\nion{$\dot{n}_{\mathrm{ion}}$}

\newcommand{\glimpse}{{\sc Glimpse}}
\newcommand{\jwst}{{\em JWST}}
\newcommand{\hst}{{\em HST}}
\newcommand{\ra}[3]{#1$^{\mathrm{h}}$#2$^{\mathrm{m}}$#3$^{\mathrm{s}}$}
\newcommand{\decl}[3]
{#1$^{\circ}$#2$'$#3$''$}

\shorttitle{GLIMPSE UVLF $z\sim7$}
\shortauthors{}

\usepackage[export]{adjustbox}
\begin{document}
\title{A GLIMPSE of the 99\%: a census of the faintest galaxies during the epoch of reionization and its implications for galaxy formation models\vspace{-1.5cm}}

\author{Hakim Atek (\includegraphics[height=1.7ex]{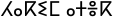})$^{1*}$, Iryna Chemerynska$^{1}$, Lukas J. Furtak $^{2,3}$, Johan Richard$^{4}$, John Chisholm$^{2,3}$, Vasily Kokorev$^{2,3}$, Michelle Jecmen$^{2,3}$, Damien Korber$^{5}$, Ryan Endsley$^{3}$, Richard Pan$^{6}$,  
Arghyadeep Basu$^{4}$, Jeremy Blaizot$^4$, Rychard Bouwens$^{7}$, Meriam Ezziati$^1$, Sylvain Heurtier$^1$, Kristen.~B.~W. McQuinn$^{8,9}$, Marcie Mun$^1$, Julian B.~Mu\~noz$^{3}$, Pascal Oesch$^{5,10}$, Joakim Rosdahl$^4$, Alberto Saldana-Lopez$^{11}$, Seiji Fujimoto$^{12,3}$, Daniel Schaerer$^{5}$}

\affiliation{$^1$Institut d'Astrophysique de Paris, CNRS, Sorbonne Universit\'e, 98bis Boulevard Arago, 75014, Paris, France}
\affiliation{$^{2}$Cosmic Frontier Center, The University of Texas at Austin, Austin, TX 78712, USA}
\affiliation{$^{3}$Department of Astronomy, The University of Texas at Austin, 2515 Speedway, Stop C1400, Austin, TX 78712, USA}
\affiliation{$^{4}$Univ Lyon, Univ Lyon1, Ens de Lyon, CNRS, CRAL UMR5574, F-69230, Saint-Genis-Laval, France}
\affiliation{$^5$Department of Astronomy, University of Geneva, Chemin Pegasi 51, 1290 Versoix, Switzerland}
\affiliation{$^6$Department of Physics and Astronomy, Tufts University, 574 Boston Ave., Medford, MA 02155, USA}
\affiliation{$^{7}$Leiden Observatory, Einsteinweg 55, NL-2333 CC Leiden, The Netherlands}
\affiliation{$^8$Space Telescope Science Institute, 3700 San Martin Drive, Baltimore, MD, 21218, USA}
\affiliation{$^9$Rutgers University, Department of Physics and Astronomy, 136 Frelinghuysen Road, Piscataway, NJ 08854, USA} 
\affiliation{$^{10}$Cosmic Dawn Center (DAWN), Niels Bohr Institute, University of Copenhagen, Jagtvej 128, K\o benhavn N, DK-2200, Denmark}
\affiliation{$^{11}$Department of Astronomy, Oskar Klein Centre, Stockholm University, 106 91 Stockholm, Sweden}
\affiliation{$^{12}$Dunlap Institute for Astronomy and Astrophysics, 50 St. George Street, Toronto, Ontario, M5S 3H4, Canada}
%\affiliation{$^{13}$Department of Astronomy, The University of Texas at Austin, Austin, TX 78712, USA}
\thanks{$^*$E-mail: \href{mailto:hakim.atek@iap.fr}{hakim.atek@iap.fr}}
\begin{abstract}

We present a comprehensive study of the galaxy UV luminosity function (UVLF) at $z=6-9$ and the resulting cosmic ionizing photon budget during the Epoch of Reionization, leveraging deep \jwst\ observations from the GLIMPSE survey. Thanks to gravitational lensing, we probe the UVLF to an unprecedented depth of $M_{\text{UV}} = -12$, approximately three magnitudes deeper than previous robust constraints. Our UVLF determination incorporates a rigorous end-to-end uncertainty framework, including statistical and systematic lensing uncertainties. We find that the $z \sim 7$ UVLF continues to rise steeply with a faint-end slope of $\alpha = -1.98_{-0.05}^{+0.06}$. Crucially, our data show no clear evidence of a turnover down to \muv $= -12.3$, with any potential curvature being significantly shallower than tentative estimates reported in previous \hst\ studies. The persistence of this faint population provides stringent constraints on galaxy formation models and cosmological simulations that predict an early flattening of the luminosity function due to radiative feedback or star-formation thresholds. Furthermore, post-JWST models specifically calibrated to match the UV-bright excess at $z > 10$ generally fail to reproduce the observed evolution toward lower redshifts and fainter magnitudes, highlighting a significant tension in our current understanding of early galaxy assembly. Integrating this UVLF with empirical constraints on the ionizing efficiency and escape fractions derived from GLIMPSE observations, we derive a comoving ionizing emissivity at $z=7$ of log(\nion / s$^{-1}$ Mpc${^-3}$) $\approx 50.85$. This high emissivity suggests that faint galaxies dominate the ionizing budget, providing enough photons to maintain reionization even in a highly clumped IGM ($C_{\text{HII}} = 5$). However, adopting higher ionizing efficiencies or escape fractions from the literature results in an early reionization history that completes by $z \approx 8$, which stands in tension with Planck CMB optical depth measurements and IGM opacity data. As our detection of faint galaxies effectively rules out a luminosity function truncation at \muv $\geq -15$, these results emphasize the need to either accurately characterize the ionizing properties of the global, low-mass galaxy population at $z > 6$, or to refine physical models of intergalactic medium clumping and its redshift evolution to maintain consistency with the observed reionization timeline.
\end{abstract}


\section{Introduction} \label{sec:intro}

In the standard $\Lambda$CDM model, galaxies grow through the accretion of gas onto dark matter (DM) halos. As gas loses energy through radiative cooling, it condenses more effectively than collisionless DM, ultimately forming stars \citep{white78,keres05,dekel09}. However, without regulatory processes, the stellar mass of galaxies would grow excessively, up to an order of magnitude beyond observed values \citep{benson03,baldry08}. Feedback from massive stars and supernova (SN) explosions mitigates this growth by heating the neutral gas, effectively suppressing star formation in low-mass systems. Additional feedback mechanisms, such as galactic winds and the reionization background, further limit the growth of small galaxies \citep[][]{gatto17,pawlik15,wu19,katz20}. Consequently, feedback strongly influences the number counts of galaxies at specific masses or luminosities, leading to a predicted flattening in the luminosity and mass functions \citep{liu16, vogelsberger20,hutter21}.

To align with observations of faint galaxies, the galaxy formation models require stellar feedback. However, such simulations rely on sub-grid physics to implement feedback prescriptions, which require empirical anchors. Such empirical constraints become challenging in this faint regime. Therefore, the shape and amplitude of the UV luminosity function (UVLF) at the faint end, along with their evolution through the epoch of reionization, serve as crucial tests for feedback models in simulations of galaxy formation.

Significant progress has been made in the past decade toward cataloging distant galaxies, largely due to the near-infrared capabilities of the \hst\ \citep[][]{bouwens15,finkelstein15}. With a sample of over a thousand sources at $z>6$, \hst\ has been instrumental in providing reliable constraints on the overall shape of the galaxy UV luminosity function. Despite these achievements, reaching the faintest galaxies, those with absolute UV luminosities below \muv $\sim -17$, remains challenging, even for the powerful \hst.

Gravitational lensing by massive galaxy clusters offers a way to extend observational limits, providing a deeper view into the reionization epoch. The Hubble Frontier Fields\footnote{\url{https://frontierfields.org/} and \url{https://archive.stsci.edu/prepds/frontier/}} (HFF) program dedicated approximately 850 orbits of optical and near-infrared imaging to six lensing clusters, marking one of the most significant investments of \hst\ time. This effort has pushed the UVLF constraints down to \muv\ = $-$15 mag, revealing that the faint-end slope of the luminosity function remains steep, with $\alpha \approx -2$ down to these luminosities \citep{atek18,bouwens21}.

In its first two years of operation, the \jwst\ has demonstrated exceptional capabilities in probing the distant Universe, largely due to its much larger light-collecting area and unprecedented sensitivity in the infrared. Hundreds of galaxies have been detected at $z>9$, reaching record-breaking distances and revealing an unexpected excess of UV-bright galaxies compared to theoretical predictions \citep[][]{castellano22,finkelstein23,atek23b,chemerynska24a,donnan23,harikane23,bouwens23,chemerynska25,naidu26}. Nearly all such studies report a steep and nearly constant faint-end slope ($\alpha \sim -2.2$ to $\alpha \sim -2.5$) and a modest evolution in the UV luminosity density across the redshift range $z=9-15$. 

However, relatively little attention has been given to the significant population of low-mass galaxies at the heart of the epoch of reionization. Deep blank-field surveys such as the NGDEEP \citep[][]{bagley24} and the JADES Origin Field \citep[JOF][]{eisenstein23}, have made important inroads by pushing to fainter magnitude limits around  $m_{5\sigma}\approx30.5$ AB. The NIRCam parallel field of the primary MIRI observations of the MIDIS program \citep{ostlin25} have pushed this depth to $m_{5\sigma}\approx30.8$ AB, \citep{perez-gonzalez23}.  Nevertheless, these programs were not able to probe the epoch of reionization, mainly due to the lack of deep observations in the blue filters such as the F090W. Notable exceptions include \citet[][]{adams24} who used a multi-field analysis to compute the UV luminosity function starting from $z=7.5$ for galaxies brighter than \muv $=-18$ mag. Therefore, after 3 years of operation, the \jwst\ has still yet to explore the fainter population of galaxies at the epoch of reionization. 

With the goal of extending the Frontier Fields into the \jwst\ era, we have carried out the \glimpse\  program, an ultra-deep NIRCam imaging survey,  particularly in the F090W band, centered on the galaxy cluster Abell S1063. In this paper, we present the first deep dive into this population by pushing to several magnitudes fainter than current \jwst\ surveys at the heart of the epoch of reionization. We provide strong constraints on the extreme faint end of the UV luminosity function at $6<z<9$ and derive important implications for state-of-the-art galaxy formation models and cosmic reionization. 

The paper is organized as follows. Section \ref{sec:observations} is dedicated to the description of observations and data reduction. The selection procedure of high-redshift galaxies is described in Section \ref{sec:selection}. In Section \ref{sec:lensing}, we briefly describe the different lensing models used in this study. The main result of the paper is presented in Section \ref{sec:uvlf}. We present the different steps involved in the construction of the UVLF, including the effective survey volume, survey completeness, and the lensing uncertainties. We discuss the implications of the faint-end slope on the galaxy formation models and simulations in Section \ref{sec:models}. We present the integrated UV luminosity and SFR density in Section \ref{sec:rhouv}. Section \ref{sec:reionization} is dedicated to the implication on the reionization history of the Universe. We provide a summary in Section \ref{sec:summary}. Throughout the paper we adopt the \citet{planck18} cosmology with H$_{0}=67.66$ km Mpc$^{-1}$ s$^{-1}$, $\Omega_{\mathrm{m}}=0.31$, and $\Omega_{\mathrm{\Lambda}}=0.69$.

\section{Observations}
\label{sec:observations}

The \jwst\ \glimpse\ survey \citep[][]{atek25} obtained NIRCam imaging of the lensing cluster Abell S1063 (RA = \ra{22}{48}{44.13}, DEC = \decl{-44}{31}{57.50}), located at a redshift of $z=0.348$. These are intrinsically the deepest observations of the distant Universe ever obtained. The imaging was captured in a total of nine filters: four short-wavelength (SW) broadbands (F090W, F115W, F150W, F200W), three long-wavelength (LW) broadbands, and two medium-band filters (F410M, F480M). The observations reach a $5\sigma$ magnitude limit of about 30.8 in the broadband filters and about 30 and 29.3 mag in the F410M and F480M medium bands, respectively. We also include ancillary data from \hst\ programs HFF \citep[][]{lotz17} and BUFFALO \citep[][]{steinhardt20} which obtained optical coverage of AS1063 and its parallel field, and the FLASHLIGHTS program (PID 15936), which provides imaging in the two optical long-pass filters F200LP and F350LP.

The data reduction process of the \glimpse\ images is described in detail in \citet{atek25}, and we only provide here a general overview. Data were reduced using a custom procedure presented in \citet[][]{endsley24}, based on the STScI pipeline with important enhancements. We employed the JWST Calibration Reference Data System (CRDS) context map {\tt jwst\_1293.pmap}. This procedure incorporates custom templates to correct artifacts and mask snowballs, 1/\emph{f} noise, as well as a 2D background subtraction step following \citet[][]{bagley23}. In order to further improve the quality of the images, we have produced flat-field calibration frames using publicly available NIRCam data obtained up to January 12, 2025. Finally, all {\tt *cal.fits} frames are processed through stage 3 of the calibration pipeline to create the final drizzled mosaics, which have a final pixel scale of 20 mas~pix$^{-1}$ for SW bands and 40 mas~pix$^{-1}$ for the LW bands. Next, we addressed contamination from the light of bright cluster galaxies (bCGs) and intracluster light (ICL). The modeling and subtraction of these sources followed the methods of \citet{ferrarese06, shipley18, weaver24a}. In the final step, we constructed an empirical model of the point spread function (PSF) using observed stars in the field. All images were matched to the broadest PSF of the F480M filter.

To ensure the inclusion of both blue and red sources, we construct two stacks of the non-PSF-convolved short-wavelength (SW) and long-wavelength (LW) images using inverse-variance weighting (IVW) maps. Source detection is carried out with \textsc{SExtractor} \citep{bertin96}, and the resulting catalogs are merged by retaining only unique sources, identified through the non-overlapping regions of their respective segmentation maps. Photometry is performed on PSF-homogenized images using \textsc{PHOTUTILS} \citep{bradley20}, adopting circular apertures with a diameter of $0\farcs{2}$. Photometric uncertainties are estimated by sampling the local background with 2,000 random apertures placed around each source and combining the resulting scatter with the Poisson noise \citep{weaver23,endsley24}. Finally, we apply aperture corrections to all fluxes using the PSF growth curve measured in the F480M filter \citep[see][]{atek25}. While such correction is well suited for point-like sources, it may introduce a photometric bias for brighter or highly amplified sources whose sizes are larger than this aperture size. To estimate such bias, we computed the difference between the aperture-corrected magnitudes in a small ($0\farcs{2}$) and a large aperture ($1\farcs{0}$) and its evolution with the observed magnitude. While a slight dependency is observed, the bias remains around 0.1 mag for sources fainter than AB=29, which remains significantly smaller than the rest of the uncertainties affecting their inferred luminosity, particularly those related to lensing magnification.

\section{Identification of high-redshift galaxies}
\label{sec:selection}

Our primary method for selecting high-redshift galaxies is based on the Lyman-break technique, which identifies the Lyman-$\alpha$ continuum break caused by absorption from the intergalactic medium. To optimize the color-color selection window for robust identification of $z>6$ galaxies and efficient rejection of contaminants, we generate mock spectral energy distributions (SEDs) using the {\sc Beagle} tool \citep[][]{chevallard16}. We model star-forming galaxy templates spanning redshifts of $z = 5$-10 with dust attenuation values ranging from $A_V = 0$ to 1 mag, adopting a Small Magellanic Cloud (SMC) extinction law \citep[][]{pei92,reddy15}. To account for possible interlopers, we include quiescent galaxy templates from the SWIRE library \citep[][]{polletta07}, based on the {\sc GRASIL} code \citep[][]{silva98}, applying dust attenuation values between $A_V = 1$ and 3 mag. Another significant source of contamination is cool stars such as M, L, and T dwarfs, which can mimic Lyman-break signatures due to their intrinsically red colors and strong absorption features. We therefore incorporate brown dwarf templates from the libraries of \citet[][]{Chabrier00} and \citet{allard01}. Synthetic photometry is then computed across all \jwst\ filters, enabling us to define optimal color-color selection criteria for isolating galaxies at $z>6$.

Based on these diagnostics, galaxy candidates satisfy the following photometric and color criteria:   
\begin{equation}
	\begin{array}{l}
		m_{090}-m_{115}>0.6\\
        \land ~ m_{090}-m_{115}>1.6+1.08~(m_{115}-m_{150})\\
	\land ~ m_{115}-m_{150}<0.8 \lor m_{150}-m_{200}<0.8 \\
    \land ~ \mathrm{SNR}_{115} > 4.0 \land ~ \mathrm{SNR}_{150} > 3.0 \land ~ \mathrm{SNR}_{200} > 3.0 \\
    \land ~ \mathrm{SNR}_{606} < 2.0 \land ~ \mathrm{SNR}_{814} < 2.0
	\end{array}
  \label{eq:dropout}
\end{equation}

\noindent where the photometric criteria ensure a significant detection in at least three broadbands and non-detection in all the bands bluewards of the Lyman break, which includes the ultra-deep \hst\ images in the F200LP and F350LP. We also reject sources that show rest-frame UV colors redder than 0.8 mag. We use a combination of two colors to avoid discarding robust sources for which the Lyman break falls within the filter and may produce red rest-UV colors. Additionally, in case of a non-detection in the dropout filter, we assign the $1-\sigma$ depth in that aperture to its magnitude to ensure that the Lyman break does not go below the color criterion. This means that the detection band is always at least 0.6 mag brighter than the detection limit in the dropout band. We also exclude all objects that have been flagged during the photometric extraction by using only objects with a flag {\tt use\_phot=1} \citep[cf.][]{atek25}. Then all selected sources are visually inspected to remove remaining artifacts, which usually are bCG subtraction residuals or contamination from bright sources.

\begin{figure}
    \centering
    \vspace{0.3cm}
    \includegraphics[width=0.9\linewidth]{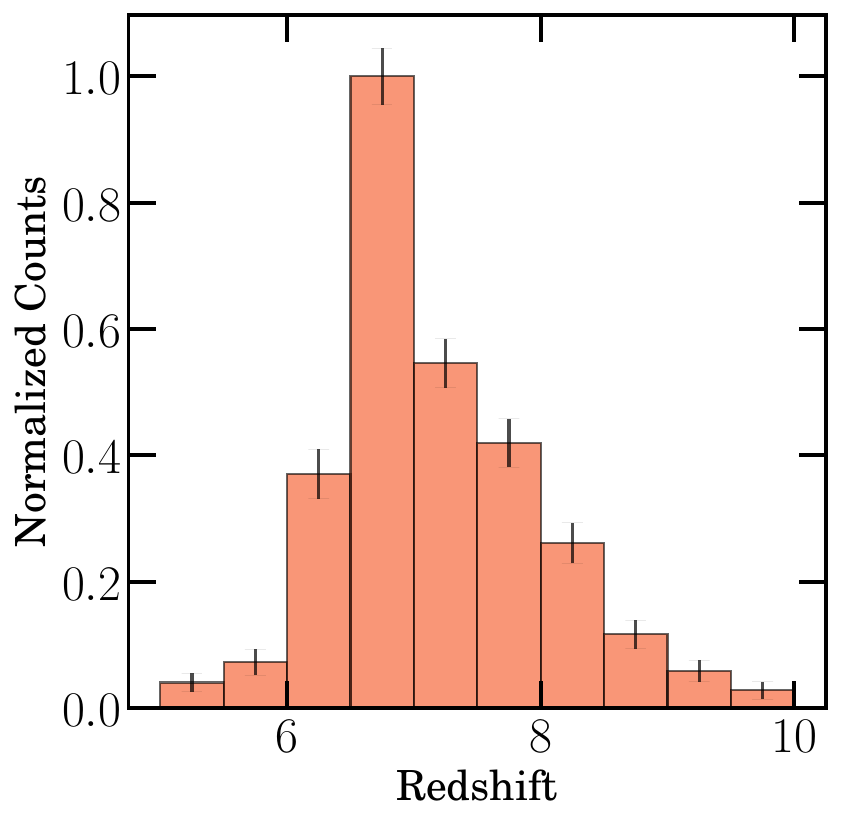}
    \caption{The redshift density distribution of the full sample normalized to unity. The distribution accounts for photometric redshift uncertainties through an MC sampling of the errors. The error bars show the dispersion in each redshift bin. While the wavelength location of the Lyman break shapes the edges of the redshift distribution, the lower number of galaxies along the high$-z$ tail is mainly due to the apparent magnitude completeness.}
    \label{fig:histo_z}
\end{figure}

A second step consists of computing photometric redshifts for this selected sample. We run the Python version of the {\sc Eazy} software \citep{Brammer2008} on the entire \glimpse\ source catalog using the 0.2\arcsec aperture photometry. We adopt the following parameters for the SED fitting procedure: a redshift grid from $z=0.01$ to $z = 30$, a  \citep{Calzetti00} dust attenuation law, an IGM prescription following \citet[][]{asada24}, and the {\tt blue\_sfhz\_13} template library. The SED fitting procedure also includes high-resolution synthetic stellar templates from the {\tt PHOENIX} library \citep{husser13}. %and brown dwarf templates \citep{burgasser14}. 
Overall, all photometric redshifts agree with the high$-z$ solution inferred by the color-color selection, meaning the high$-z$ solution is preferred to the low$-z$ one, with the exception of a few catastrophic cases for which we forced a high-redshift solution during the fitting run. For the rest of our calculations, we adopt the {\sc Eazy} best-fit photo$-z$ as the source redshift with the associated uncertainties. In the end, our final sample contains 276 sources with photometric redshifts spanning the range $z\sim6$ to $z\sim9$ (cf. Fig. \ref{fig:histo_z}).

\begin{figure}
    \centering
    \includegraphics[width=0.9\linewidth]{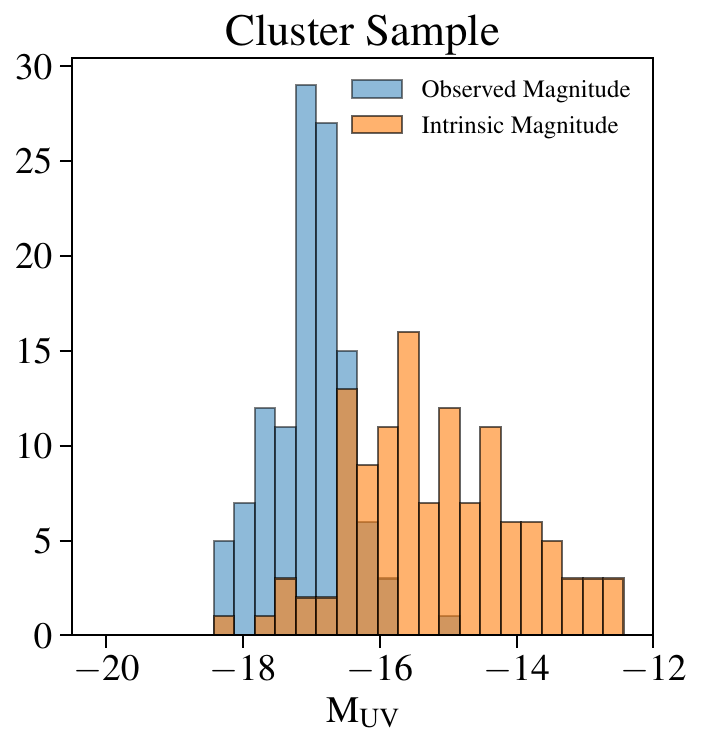}\\
     \includegraphics[width=0.9\linewidth]{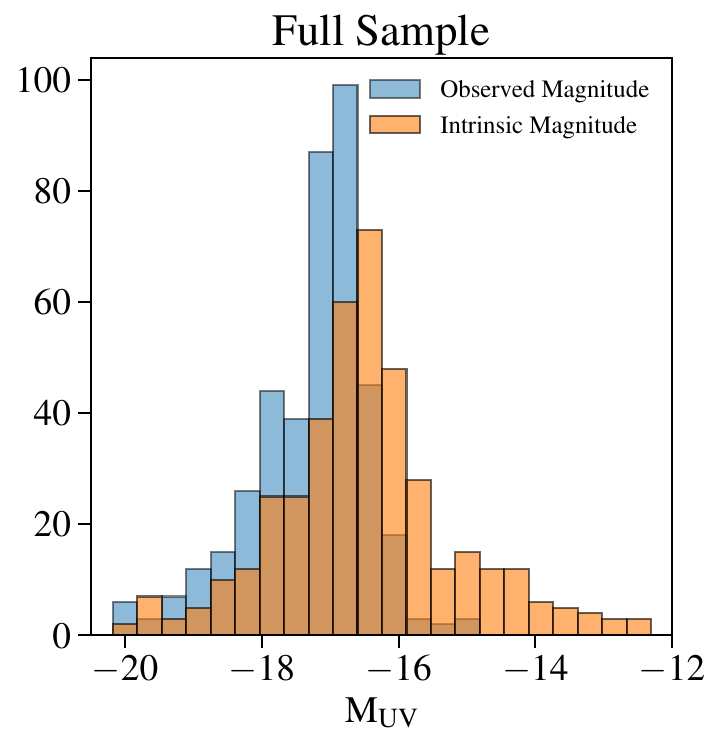}
    \caption{The absolute magnitude distribution of the galaxy sample at $6<z<9$. The {\it top panel} shows the distribution for the NIRCam module centered on the lensing cluster, while the {\it bottom panel} is for the entire field of view. The blue histogram represents the observed magnitude, while the orange histogram shows the intrinsic (de-lensed) magnitude.}
    \label{fig:muv}
\end{figure}

\section{Lensing Models}
\label{sec:lensing}

As part of the Hubble Frontier Fields (HFF) program \citep[][]{lotz17}, AS1063 is a well-studied galaxy cluster that has been the subject of multiple gravitational lensing mass models developed by different teams within the HFF community \citep[][]{richard14a,monna14,zitrin2015,caminha16,bergamini2019}. Building on previous HFF constraints, primarily based on the identification of multiple-image systems in \hst\ and MUSE observations, and incorporating newly identified multiple images from the \glimpse\ NIRCam imaging, we have constructed a new lensing model (Furtak et al. in prep.). In particular, this model significantly improves the lensing constraints in the north-eastern sub-halo, which was poorly constrained by the HFF data alone. The model achieves an uncertainty of {\tt rms}=0.32\arcsec in the predicted position of lensed galaxies \citep{furtak26}. This parametric model is built using the fully analytic code of \citep[][]{zitrin2015}, previously used to model multiple clusters \citep[e.g.][]{pascale2022,furtak23b}. The model uses a total of 75 multiple images, among which 24 have spectroscopic redshifts derived from MUSE and \hst\ data \cite[see][for more details]{atek25}. 

In addition to this fiducial model, we use two independent lens models in the present study in order to assess the systematic uncertainties and their impact on the computed luminosity function. The second model presented in \citet[][]{Beauchesne24} is constructed using a joint, self-consistent approach that combines strong gravitational lensing constraints with X-ray observations of the intra-cluster gas and galaxy kinematics. The cluster mass distribution is decomposed into its main physical components: large-scale dark matter halos, the X-ray emitting intra-cluster gas, and galaxy-scale perturbations associated with cluster members. Strong-lensing constraints from multiple-image systems are combined with X-ray surface-brightness measurements to directly constrain the gas mass component, while spectroscopic measurements of cluster-member velocity dispersions are used to calibrate the galaxy mass scaling relations. All observables are incorporated within a single Bayesian framework, allowing the parameters of the different mass components to be optimized simultaneously and consistently. 

The third model \citep[][]{richard14} is based on a parametric reconstruction using {\tt Lenstool} \citep[][]{kneib96,jullo07}. The mass distribution of the cluster is modeled using a combination of strong-lensing and weak-lensing constraints. Similarly to the first model, the strong-lensing component is constrained by spectroscopically confirmed multiple-image systems identified in deep \hst\ imaging and MUSE observations. The model was developed to reproduce the observed positions and redshifts of the multiply imaged background sources and provides a lensing model over the cluster core. The high density of secure spectroscopic constraints from MUSE enables a robust determination of the mass distribution in the central region of the cluster.

\section{The Galaxy UV Luminosity Function at $6<z<9$}
\label{sec:uvlf}

Following \citet[][]{atek18,chemerynska25}, we compute the UV luminosity function using a volume-based determination by working directly in the source plane. To derive the UVLF, the observed galaxies are grouped into bins of absolute UV magnitude spanning the range from $M_{\rm UV} = -19$ to $-12$, adopting a uniform bin size of $\Delta M = 1$ mag. This binning scheme provides a compromise between sufficient number statistics in each bin and adequate resolution to trace the shape of the luminosity distribution, particularly toward the faint end.

For each redshift interval, the UVLF is estimated by counting the number of detected sources within each magnitude bin and normalizing this quantity by the corresponding survey volume over which such galaxies could be observed. The luminosity function is therefore defined as
\begin{equation}
\phi(M_i)\,\mathrm{d}M_i = \frac{N_{{\rm obj},i}}{V_{\rm eff}(M_i)} ,
\end{equation}
where $N_{{\rm obj},i}$ denotes the number of galaxies whose absolute UV magnitudes fall within the $i^{\rm th}$ bin centered on $M_i$. The effective comoving volume, $V_{\rm eff}(M_i)$, accounts for the redshift interval considered, the survey area, and observational selection effects such as detection completeness. The absolute magnitude, both observed and intrinsic (de-lensed), distribution of the sample is shown in Figure \ref{fig:muv}.

\subsection{Survey Completeness}
\label{sec:comp}

This analysis adopts a source-plane-based methodology, similar to the approach introduced in \citet[][]{atek18} and \citet[][]{chemerynska25}. In this framework, mock galaxies covering a broad range of intrinsic physical properties and redshifts are inserted directly into the reconstructed source plane associated with the survey field, as defined by the adopted cluster mass model. By performing the simulations in the source plane, all gravitational lensing effects are naturally incorporated in a self-consistent manner. These include flux magnification, morphological distortion, geometric deflection, and the production of multiple images. As a result, this technique captures the complex impact of lensing on source detectability, including spatially dependent variations in completeness. This provides a robust way to quantify selection effects and observational biases and systematic uncertainties induced by strong lensing when deriving intrinsic galaxy population statistics.

In total, we generate 150,000 simulated galaxies in the source plane, covering the redshift interval $5 \leq z \leq 12$ and absolute UV magnitudes in the range \muv$=-20$ to $-10$ mag. To better probe the regime where detection completeness declines rapidly, 50,000 of these mock galaxies are assigned magnitudes fainter than \muv$=-15$. To adequately sample regions of high gravitational magnification in the source plane, which map onto strongly stretched areas in the image plane, we increase the local density of simulated sources by a factor of ten in these regions. This strategy ensures a reliable characterization of completeness where lensing effects are strongest. A key ingredient in the completeness estimates is the assumed size distribution of the simulated galaxies \citep[][]{bouwens17a,atek18}. At fixed intrinsic luminosity, galaxies with larger physical sizes, especially when combined with significant lensing-induced distortion, tend to have lower surface brightness and are therefore more difficult to detect. We account for this effect by adopting empirical size-luminosity relations measured at comparable redshifts: the relation from \cite{shibuya15} is used for relatively bright galaxies (\muv$<-16.5$ mag), while the relation from \cite{yang2022} is applied to fainter sources. We draw a log-normal distribution with a mean half-light radius following these relations \citep[e.g.,][]{grazian12, huang13}.

The next step is to derive synthetic photometry for the simulated galaxies based on spectral energy distribution (SED) models generated with {\tt BEAGLE} \citep{chevallard16}. The templates rely on the stellar population synthesis models of \citet[][]{bc03}, assuming a constant star-formation history. The ionization parameter is allowed to vary over $\log(U)=-4$ to $-1$, and dust attenuation is modeled using an SMC-type extinction curve \citep{pei92}. A fixed subsolar metallicity of $Z=0.1\,Z_{\odot}$ is adopted. We assigned a visual attenuation $A_{\rm V}$ to galaxies as a function of their \muv\ following the relationship derived by \citet{topping24a}. For galaxies fainter than \muv$=-16$ the attenuation becomes negligible.

\begin{figure}
    \centering
    \includegraphics[width=\linewidth]{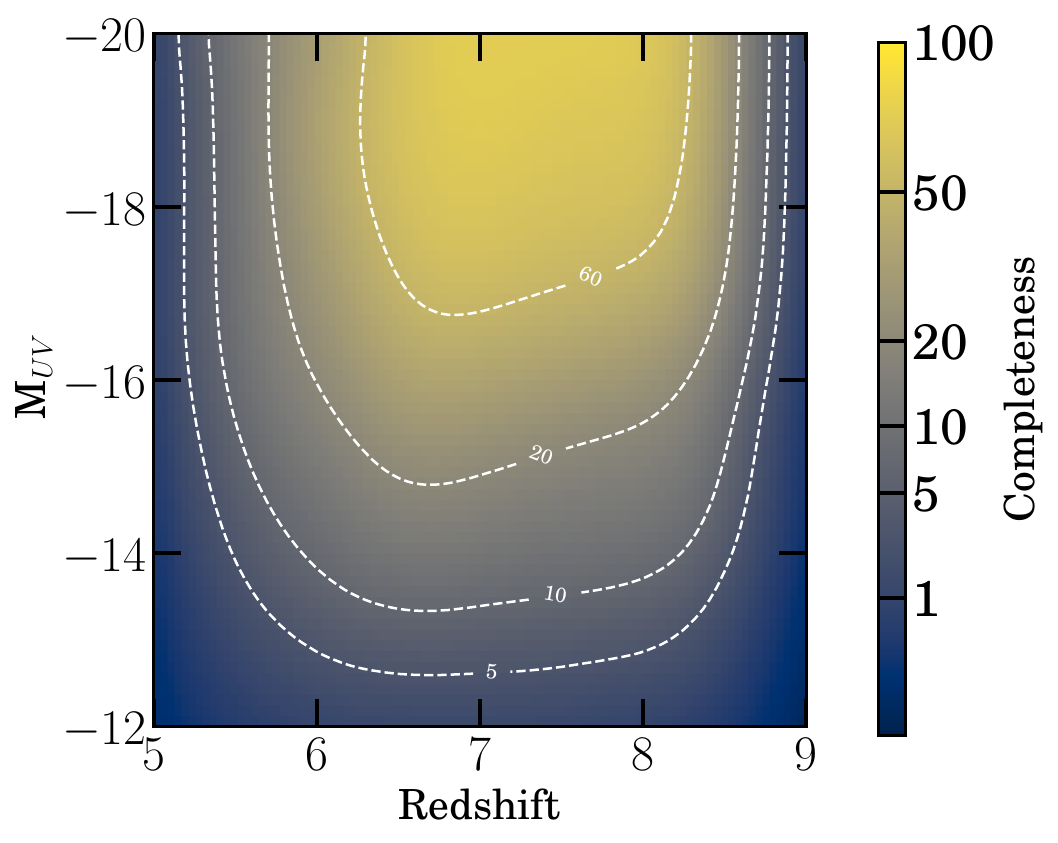}
    \caption{The 2D completeness map of the galaxy selection procedure as a function of both the intrinsic absolute magnitude \muv\ and the redshift (see Section \ref{sec:comp} for details). }
    \label{fig:comp}
\end{figure}

These model SEDs are redshifted to the previously assigned redshift and scaled to reproduce the assigned mock (observed) magnitudes, including magnification, in the detection band (F115W), which probe the rest-frame UV. Based on these fluxes and the half-light radii, galaxy images are generated with {\tt GalSim} \citep{rowe2015galsim}. We adopt S\'ersic surface-brightness profiles, parameterized primarily by the total flux and half-light radius discussed above. The simulated sources are subsequently processed through the cluster lensing model, incorporating both magnification and shear, and finally convolved with the empirical NIRCam PSF. To avoid crowding effects , we randomly distribute 100 of these galaxies at a time in a single real NIRCam image. The completeness function is calculated by comparing a selected sample based on the same extraction and color-color criteria as in observations to the input mock catalog. We remove multiple images from the selected galaxies by keeping the brightest source only. Figure \ref{fig:comp} shows the final 2D completeness map as a function of redshift and UV magnitude.  

\subsection{Effective Survey Volume}

\begin{figure*}
    \centering
     \includegraphics[width=0.47\linewidth]{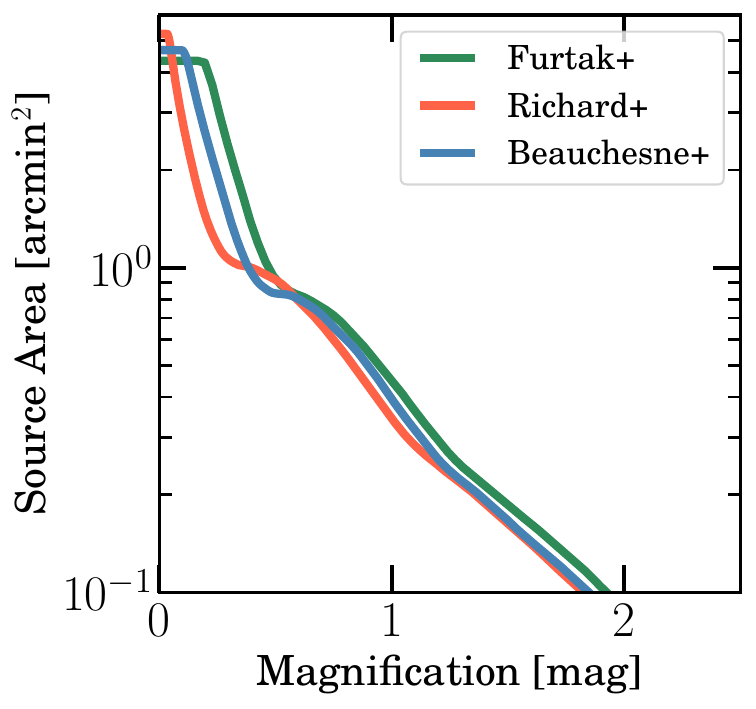}
    \includegraphics[width=0.44\linewidth]{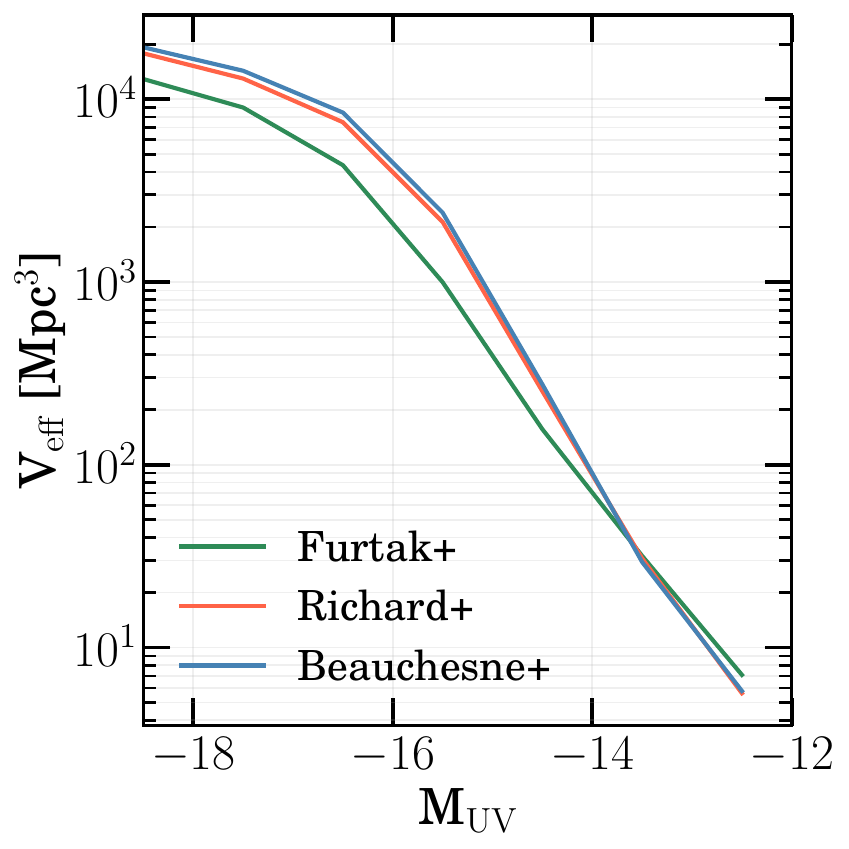}
    \caption{Survey volume characterization for the AS1063 cluster field based on three independent lensing mass models: Furtak et al. (in prep) \citet{Beauchesne24} and \citet{richard14} {\bf Left panel:} Cumulative effective source plane area ($arcmin^{2}$) as a function of magnification ($\mu$, expressed in magnitudes) for the three mass models. The differences in area highlight the systematic variations in the predicted lensing power of each model. {\bf Right panel:} The resulting effective survey volume ($V_{\text{eff}}$) in $Mpc^{3}$ as a function of intrinsic absolute magnitude ($M_{\text{UV}}$). These curves incorporate the magnification-dependent source area from the left panel as well as image-plane completeness functions, providing the final normalization used to derive the UVLF shown in Figure \ref{fig:lf7}.    
    }
    \label{fig:eff_vol}
\end{figure*}

The effective comoving volume is governed by the source-plane area accessible at a given magnification and redshift. To quantify this, we evaluate the area corresponding to each magnification value from the lens model at the given redshift, together with the minimum magnification required for a galaxy of a given intrinsic magnitude to be detectable within the survey.

The accessible surface area is obtained by integrating over the area where the magnification exceeds this detection threshold. In practice, the total projected area of the cluster sets the upper bound on the effective volume, while regions of very high magnification contribute progressively less area. Consequently, although strong lensing enhances the detectability of intrinsically faint galaxies, the corresponding source-plane area and thus the associated comoving volume decreases substantially at the highest magnifications. The reduction of the effective area as a function of magnification is already folded into the completeness simulations since this is performed directly in the source plane. Therefore, to obtain the effective survey volume, we simply multiply the total survey volume by the completeness value at this intrinsic magnitude. 

Figure \ref{fig:eff_vol} shows the resulting effective survey volume marginalized over the intrinsic absolute magnitude.

\subsection{UVLF Uncertainties}
\label{sec:uncertainties}

While computing the UVLF we also assess and incorporate various errors. Uncertainties affecting individual intrinsic UV magnitudes are propagated into the luminosity-function estimate. These include errors in magnification factors, photometric flux measurements, and photometric redshift determinations. To incorporate these effects, we perform Monte Carlo realizations in which galaxy magnitudes are randomly perturbed according to their estimated uncertainties, allowing objects to migrate between magnitude bins. This procedure mitigates biases linked to measurement errors and reduces sensitivity to the specific placement and width of the adopted bins.

In addition to the statistical lensing uncertainties that are propagated through the computation of the UVLF, we also include the lensing systematics by using three independent models for the whole procedure, which impacts both the magnification, i.e. the intrinsic magnitude, and the effective volume. Figure \ref{fig:eff_vol} shows the derived survey volume for each lensing model. 

Beyond the uncertainty associated with the effective survey volume, additional sources of error are considered when estimating galaxy number densities in each luminosity-function bin. We first account for statistical fluctuations arising from finite galaxy counts by adopting the confidence limits derived by \cite{Gehrels86}, which are appropriate for samples where small-number statistics dominate. We further include the expected field-to-field variation in galaxy abundance due to large-scale structure. The fractional uncertainty from cosmic variance is computed for each redshift and magnitude interval using the lensing-corrected survey volume together with the calculator presented by \cite{Trapp20}. This analysis indicates typical cosmic-variance uncertainties in the range ${\rm CV}\approx 10-30\%$.

\begin{figure}[!ht]
    \centering
    \includegraphics[width=\linewidth]{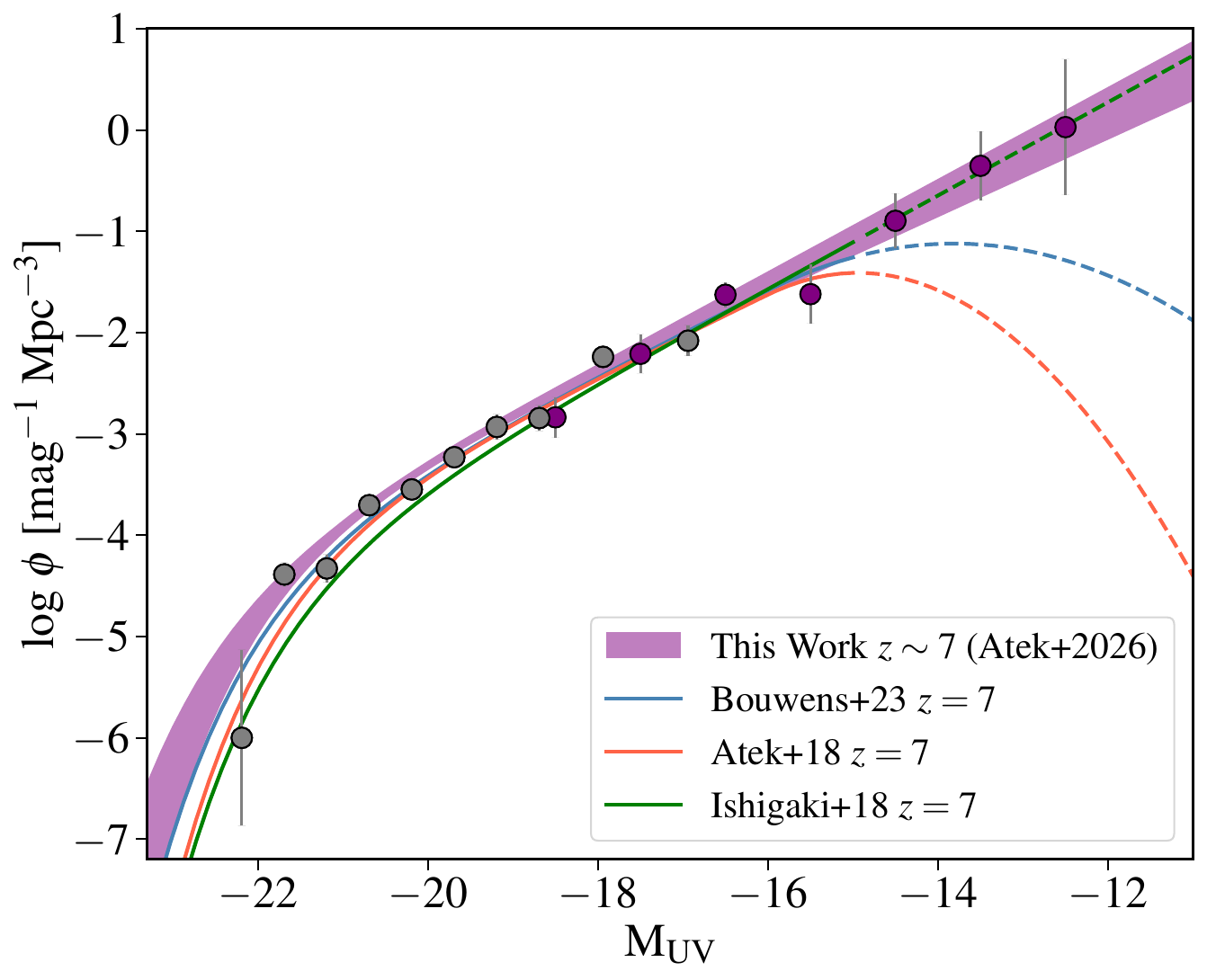}
    \caption{The UV luminosity function at $z\sim7$. The present determinations are shown by purple points, while literature points from \citet[][]{bouwens21} are denoted by gray circles. The purple curve shows $1-\sigma$ confidence interval of the best-fit Schechter function to the combined data points. The blue, red, and green curves show literature best-fit LFs from \citet[][]{bouwens22c}, \citet[][]{atek18}, and \citet[][]{ishigaki18}. Solid lines mark the robust regime for these LF determinations, while the dashed lines denote uncertain extrapolations. 
    }
    \label{fig:lf7}
\end{figure}

\subsection{Constraints on the extreme faint end of the UV luminosity function at z$>$6}

The final UV luminosity function at $z\sim7$ is shown in Figure~\ref{fig:lf7}, and the corresponding binned LF measurements are listed in Table~\ref{tab:binned_lf}. A key result of this study is the new constraints on the faint-end slope, which now reach magnitudes as faint as $M_{\text{UV}} = -12$. This limit is approximately three magnitudes deeper than the robust constraints previously derived from the Hubble Frontier Fields (HFF) program. The best fit is derived using a Markov Chain Monte Carlo procedure, including the various uncertainties discussed in Section \ref{sec:uncertainties}. The data points combine our LF measurements (purple points) and those of \citet[][]{bouwens21} which better sample the bright end of the LF (gray points). 

\begin{figure}
    \centering
    \includegraphics[width=\linewidth]{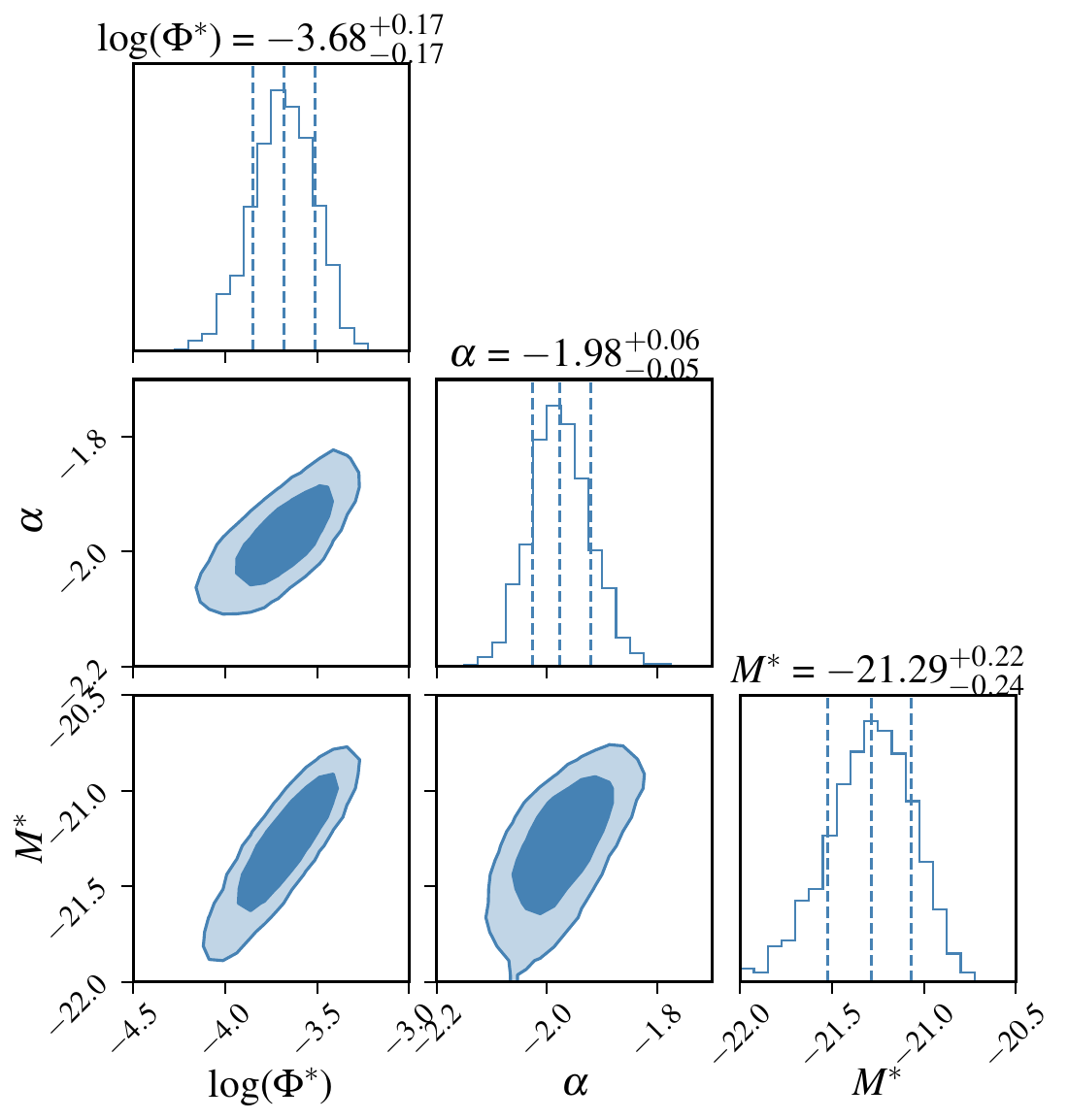}
    \caption{Posterior probability distributions for the Schechter function parameters ($\log \Phi^*$, $\alpha$, $M^*$) describing the $z \approx 7$ UVLF. This MCMC analysis is based on the final binned data, which accounts for both observational errors and the systematic variance between the three lensing mass models (cf. Figure \ref{fig:lf_lens_models}). The diagonal panels display the marginalized one-dimensional histograms for each parameter, with vertical dashed lines indicating the median and the 16th/84th percentile confidence intervals. The resulting best-fit parameters correspond to the UVLF and confidence intervals presented in Figure \ref{fig:lf7}.
    }
    \label{fig:corner}
\end{figure}

\begin{figure}
    \centering
    \includegraphics[width=\linewidth]{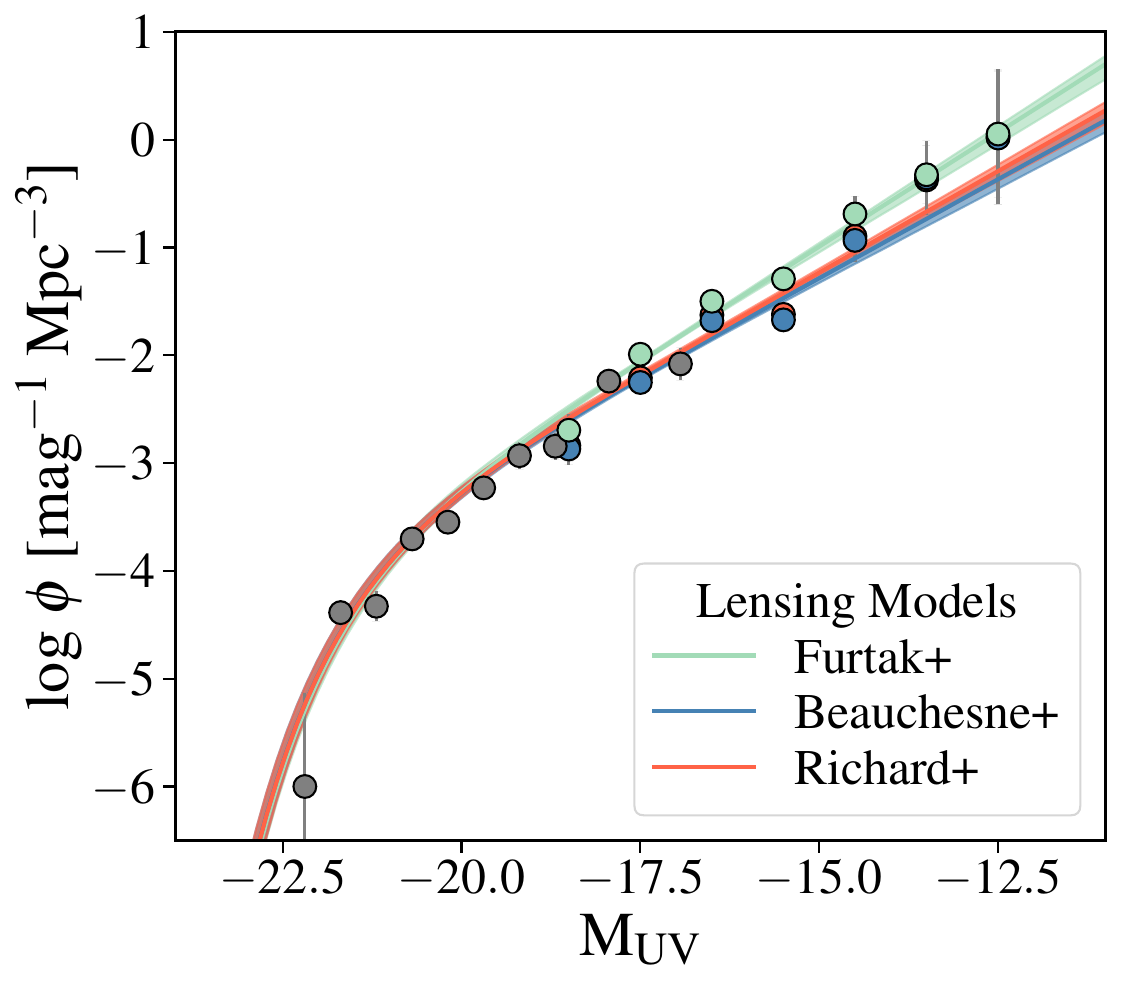}
    \caption{The rest-frame UVLF at $z \approx 7$ derived using three independent gravitational lensing mass models to evaluate systematic uncertainties (see Section \ref{sec:lensing}). The data points and corresponding Schechter function fits represent results based on the mass modeling of Furtak et al. (in prep) \citet{Beauchesne24} and \citep{richard14}. Each model derivation self-consistently incorporates lensing-related corrections, including individual source magnification factors, completeness functions, and the effective source-plane area as a function of magnification. The divergence observed at all magnitudes highlights the impact of mass-model systematics on the determination of the faint-end slope $\alpha$.
    }
    \label{fig:lf_lens_models}
\end{figure}

The luminosity function is typically parameterized using the Schechter form \citep{schechter76}. Expressed in terms of UV magnitudes, the number density $\Phi(M)$ is:

\begin{equation}
    \Phi(M) = 0.4 \ln(10) \Phi^{*} 10^{0.4(M^{*} - M)\alpha + 1} \exp(-10^{0.4(M^{*} - M)}),
    \label{eq:schechter}
\end{equation}

where $M^{*}$ is the characteristic magnitude, $\Phi^{*}$ is the normalization density, and $\alpha$ is the faint-end slope. The best-fit faint-end slope remains steep at $\alpha = -1.98 _{-0.05}^{+0.06}$, with no clear evidence of a turnover. Although some previous HFF determinations in lensed fields suggested tentative hints of a turnover around $M_{\text{UV}} \approx -14$ \citep[e.g.,][]{atek18,bouwens22c}, those findings remained consistent with a steep Schechter-like faint-end slope within the uncertainties. In particular, \citet{atek18} have shown that extrapolations beyond $M_{\text{UV}} \approx -15$ were largely unreliable; these regimes are denoted by dashed lines in Figure~\ref{fig:lf7} to contrast with the high-confidence constraints presented with solid lines. The remaining best-fit Schechter parameters are a characteristic magnitude $M^{*}= -21.29 _{-0.24}^{+0.22}$ and log($\Phi^{*}$)$=3.68 _{-0.17}^{+0.17}$. The corner plot showing the posterior distribution of all Schechter parameters is presented in Figure \ref{fig:corner}. Following the UVLF uncertainties discussed in Section \ref{sec:uncertainties}, Figure \ref{fig:lf_lens_models} shows different UVLFs computed independently using the three different lensing models. When restricting the analysis to the high-completeness regime above $\sim 20\%$, which corresponds to absolute magnitude brighter then \muv$=-15$, we find LF parameters in broad agreement with previous results with $\alpha = -2.05 _{-0.04}^{+0.05}$, $M^{*}= -21.02 _{-0.04}^{+0.04}$, and log($\Phi^{*}$)$=3.49 _{-0.04}^{+0.07}$.

\begin{figure}
    \centering
    \includegraphics[width=\linewidth]{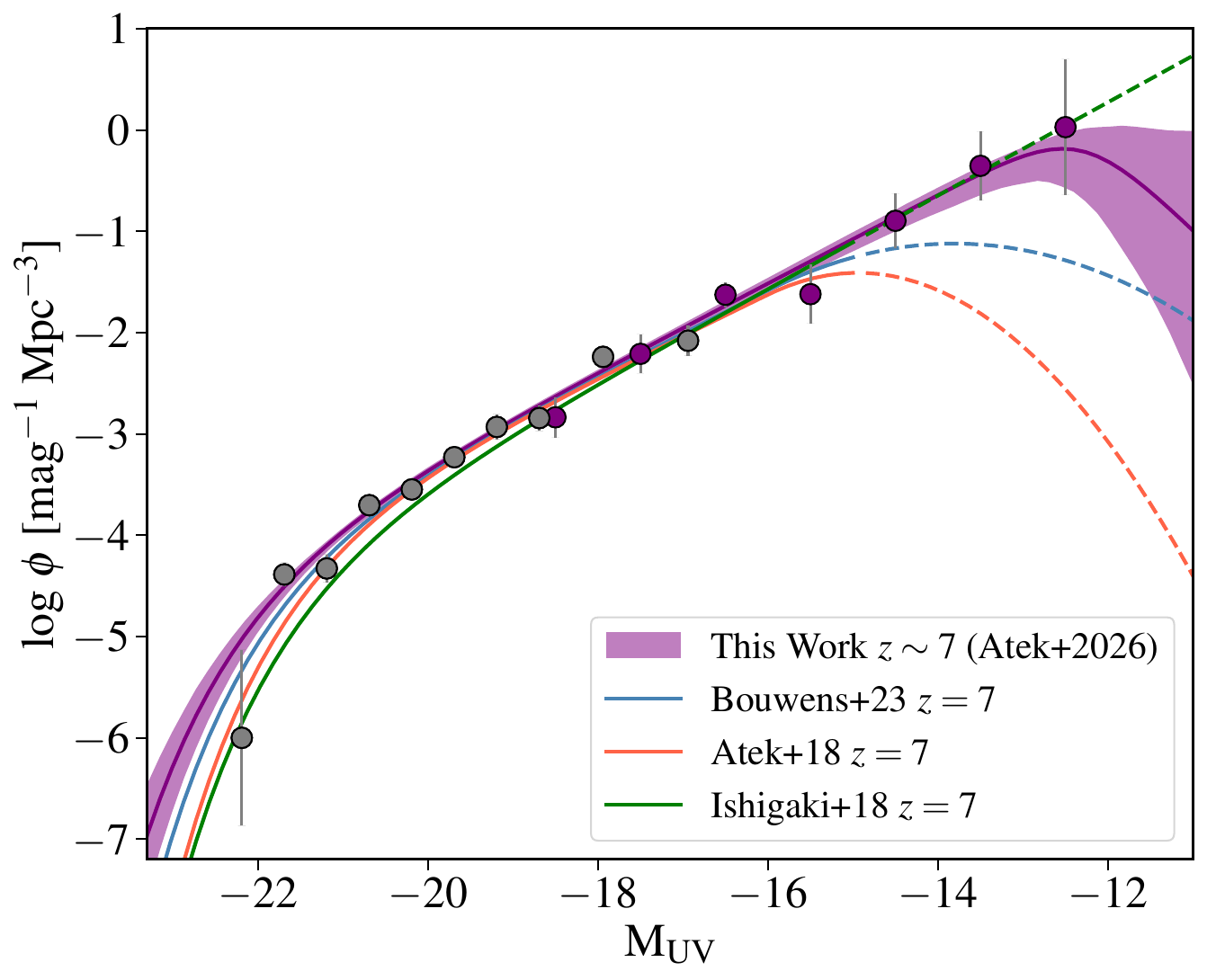}
    \caption{Same as Figure \ref{fig:lf7} but using a modified Schechter function allowing a rollover of the UVLF at the faint end. }
    \label{fig:lf_turn}
\end{figure}

\begin{figure}
    \centering
    \includegraphics[width=\linewidth]{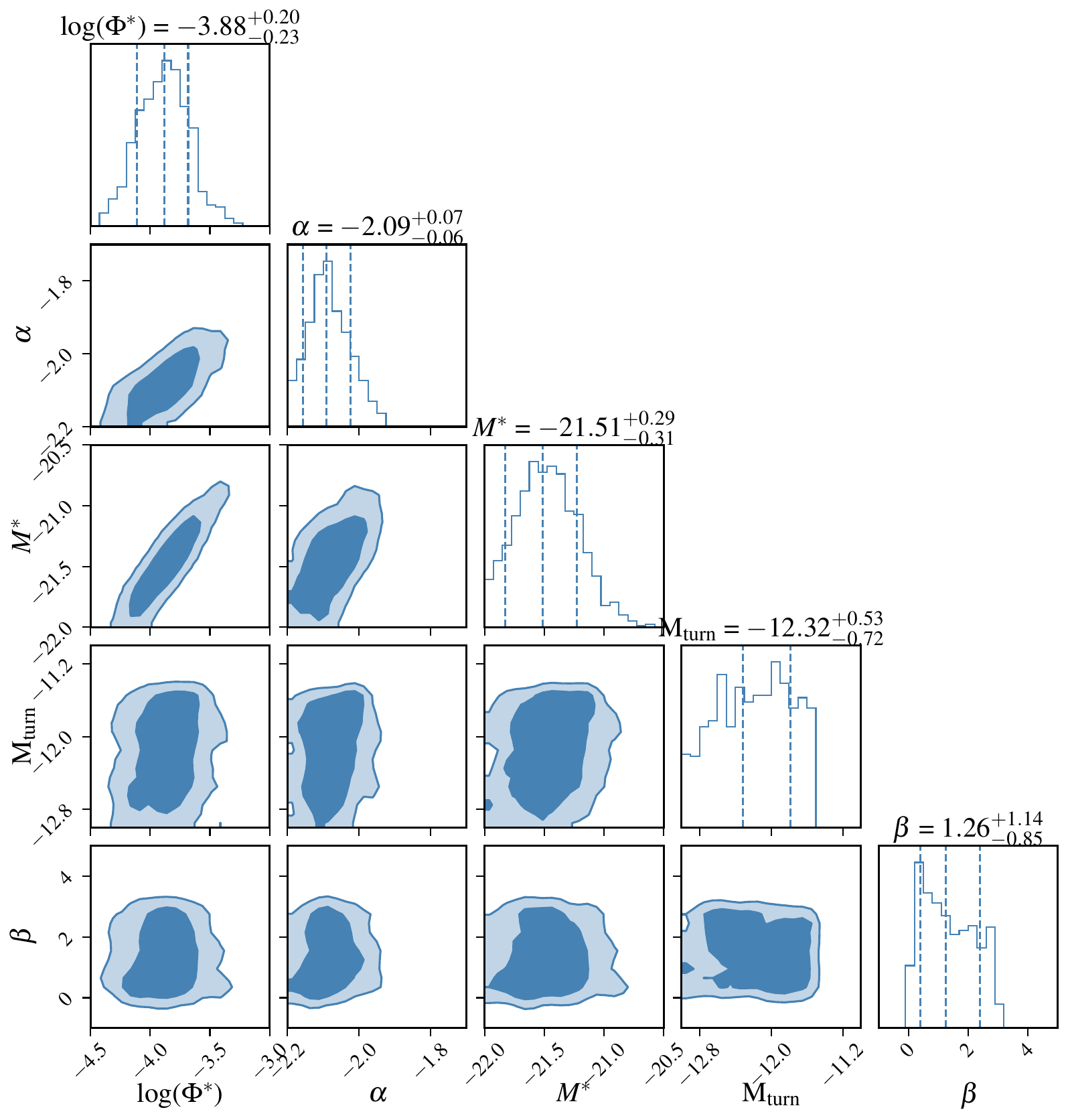}
    \caption{Same as Figure \ref{fig:corner} but using a modified Schechter function allowing a rollover of the UVLF at the faint end, providing two additional parameters, the turnover magnitude M$_{\rm turn}$ and the turnover curvature $\beta$.
    }
    \label{fig:corner_turn}
\end{figure}

Given the relatively large uncertainties in the faintest bin of the luminosity function, we explore the possibility of a turnover at magnitudes fainter than $M_{\text{UV}} = -13$. Following the methodology of the HFF studies \citep[e.g.,][]{bouwens17b,atek18}, we modify the standard Schechter function below $M_{\text{UV}} = -16$ by applying a curvature term: $10^{-0.4 \beta (M + 16)^{2}}$, where $\beta$ represents the turnover parameter. The best-fit relation yields an even steeper faint-end slope of $\alpha = -2.07 _{-0.06}^{+0.07}$, a curvature parameter of $\beta = 1.26 _{-0.85}^{+1.14}$, and a turnover magnitude of $M_{\mathrm T} > -12.3$ (at 1$\sigma$ level, see Figure \ref{fig:lf_turn}). While these results carry larger uncertainties than the pure Schechter model, they indicate a shallower curvature than previously reported and, crucially, suggest that the LF turnover, if any, likely occurs at magnitudes fainter than $M_{\text{UV}} = -13$. The posterior distributions for the modified Schechter parameters are shown in Figure \ref{fig:corner_turn}. Other studies exploring the faint end of the UVLF based on the HFF data include \citet{yue18}, who find a lower limit to the turnover magnitude at $M_{\mathrm T} > -14.6$. Similarly, \citet{livermore17} find a steep faint-end slope of $\alpha = -2.06 \pm 0.04$ at $z=7$, with no evidence of a turnover before $M_{\text{UV}} \approx -13.5$.
Again, when restricting the analysis to absolute magnitude brighter than \muv$=-15$, we find the following LF parameters $\alpha = -2.07 _{-0.03}^{+0.04}$, $M^{*}= -21.28 _{-0.41}^{+0.20}$, log($\Phi^{*}$)$=3.69 _{-0.15}^{+0.17}$, and $M_{\mathrm T} > -13.9$ at 1$\sigma$ level.

\section{Implications for galaxy formation models}
\label{sec:models}

\begin{figure*}
    \centering
    \includegraphics[width=0.48\linewidth]{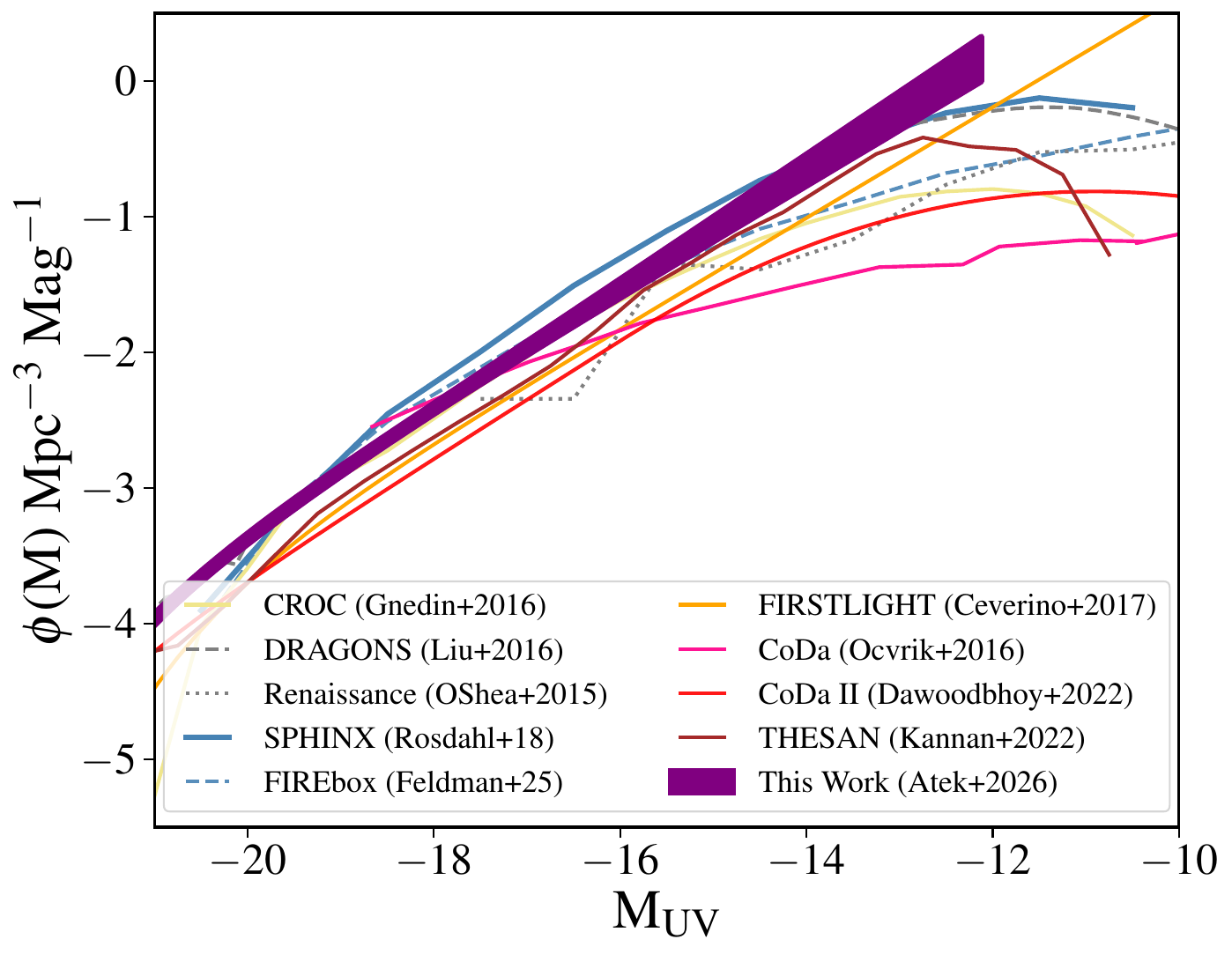}
        \includegraphics[width=0.48\linewidth]{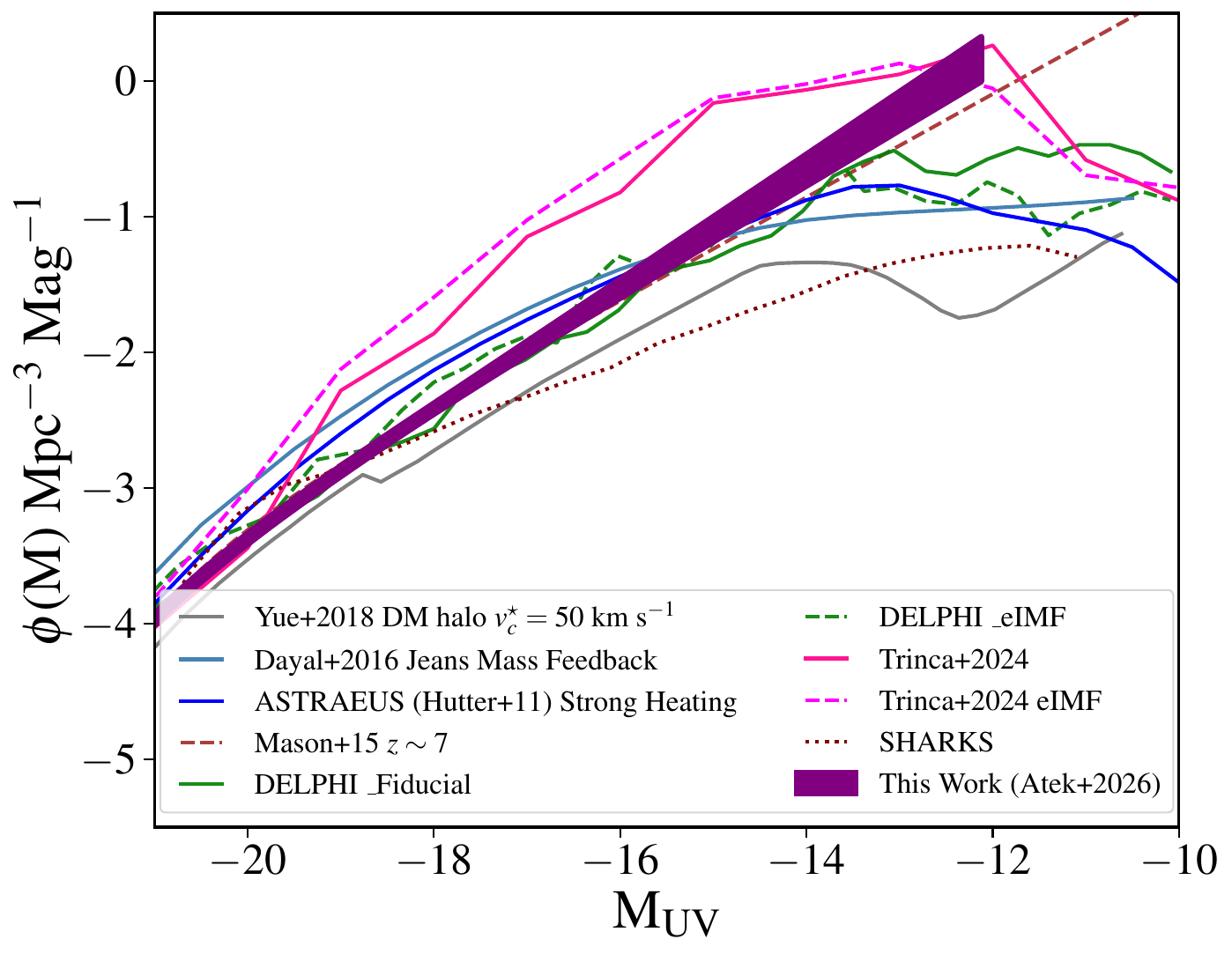}
    \caption{Constraints on the galaxy formation models from the faint-end of the UV luminosity function at $z \sim 7$. In both panels, the purple-shaded region represents our results compared to theoretical predictions. {\bf Left panel:} Comparison with cosmological hydrodynamical simulations at similar redshifts: {\sc CROC} \citep{gnedin16}, {\sc DRAGONS} \citep{liu16}, {\sc ENZO} \citep{oshea15}, {\sc Sphinx} \citep{rosdahl18}, {\sc CoDa} \citep{ocvirk16,ocvirk20,dawoodbhoy23}, {\sc THESAN} \citep[][]{kannan22}, {\sc FIRSTLIGHT} \citep{ceverino17}, {\sc FIREBox} \citep{feldmann23}. {\bf Right panel:} Comparison with pre-JWST semi-analytical models: {\sc Astraeus} \citep{hutter21}, {\sc Delphi} \citep{mauerhofer25}, {\sc CAT} \citep{Trinca24}, {\sc SHARKS} (Lagos et al. in prep), \citet{mason15,dayal15,finlator17,yue18}. 
    }
    \label{fig:lf_models}
\end{figure*}

As illustrated in Figure~\ref{fig:lf_models}, we compare our observed luminosity function with a wide suite of cosmological simulations and semi-analytical models. While our results show a consistent rise in the number density down to $M_{\text{UV}} \approx -12$, current theoretical models exhibit significant divergence in both normalization and the presence of a faint-end turnover.

\subsection{Comparison with Cosmological Simulations}

The observed disagreements primarily stem from the varying implementations of feedback and numerical resolution in simulations:

\subsubsection{Radiative and SN Feedback} The {\sc Sphinx} simulations \citep{rosdahl18} align more closely with our observed steep faint-end slope, albeit with a flattening beyond \muv=$-13$. Using high-resolution radiation hydrodynamics, these simulations achieve a physical resolution of approximately 10 parsecs and a star-particle mass resolution of $10^3 M_{\odot}$. This level of precision allows for a detailed treatment of the interstellar medium, capturing the complex interplay between radiation and gas at the scales where individual star clusters form and interact with their surroundings. Because internal feedback mechanisms take precedence over global radiative heating in these simulations, small galaxies continue forming stars even in the presence of a UV background, mirroring the continuous rise in galaxy counts down to \muv $\approx -13$ seen in current observations. In the case of a steep faint end rather than a turnover at \muv $\approx -12$, the feedback prescriptions, specifically the calibrated strength of radiative and supernova feedback, may still require fine tuning for the most dwarf-like systems.
    
The {\sc CROC} (Cosmic Reionization On Computers) simulations \citep{gnedin16} model the epoch of reionization across large statistical volumes. Unlike higher-resolution suites, {\sc CROC} operates at a coarser physical resolution of $\sim 100$ pc, which is insufficient to resolve the internal structure of the interstellar medium or individual star-forming regions. To compensate for this, {\sc CROC}, like numerous simulations, relies on sub-grid recipes for star formation and feedback. These recipes are calibrated against the UVLFs to ensure the simulation produces a realistic number of galaxies, just like the SN feedback in the {\sc Sphinx} simulations is calibrated to reproduce pre-\jwst\ $z\sim6$ UVLF. However, because these calibrations are based on the brighter galaxies, the model's physics remains largely untested in the extremely faint galaxy regime (\muv $> -13$). If the model's star formation efficiency is too low for a given halo mass, it will under-predict the number density ($\phi$) at such observed magnitudes. 

The {\sc ENZO}-based {\sc Renaissance} simulations \citep{oshea15} reach a physical resolution of approximately $19$ pc, and a star-particle mass resolution of $2.9\times10^{4}$\msol. The simulations focus on the transition from Population III stars to the first galaxies. By simulating three distinct cosmic environments, a high-density volume, an average region, and a void, this suite provides a look at how local density affects the onset of star formation. Similarly to {\sc CROC}, we also observe a flattening at the faint end of the LF which directly results from a decreasing star formation efficiency in the lowest-mass halos because of radiative feedback and less efficient gas cooling. Since the simulations correspond to a redshift of $z=12$, an increase in the overall normalization is expected toward $z=7$.

The {\sc FIREbox-HR} simulation \citep{feldmann25} is a larger volume simulation that is part of the {\sc FIRE} (Feedback In Realistic Environments) project with a baryonic mass resolution of $7800 M_{\odot}$ targeting galaxy formation in the high-redshift Universe. {\sc FIREbox-HR} is a re-run of the {\sc FIREbox} volume \citep{feldmann23} and resolves the multi-phase structure of the ISM and the resulting multi-channel stellar feedback. While other {\sc FIRE} simulations often focus on a single zoom-in, {\sc FIREbox-HR} is a representative volume of $22 \text{ cMpc}$ a side that applies the {\sc FIRE} feedback recipe, which emphasizes bursty star formation histories. The simulation uses a uniform UV background. The observed flattening of the UVLF below \muv=$-15$ is the result of  star formation suppression in tiny galaxies due to the mechanical feedback and the UV background heating.

The {\sc FirstLight} project \citep{ceverino17} is a suite of zoom-in cosmological simulations designed to follow the formation of thousands of individual galaxies during the epoch of reionization. It achieves a stellar mass resolution of $100$\msol and a physical resolution of $10$ to $20$ parsecs, comparable to the {\sc Sphinx} simulations.  It includes SN feedback, photo-ionization heating, and radiation pressure. Instead of full radiative transfer, {\sc FirstLight} includes a simple model of gas self-shielding, where the UV background is reduced for a given gas density threshold. While the slope of the UVLF is shallower than our observations, the high mass resolution allows to predict a continuous rise in the number density of faint galaxies down to the observational limit, albeit with a lower normalization likely due to the slight redshift evolution between $z=8$ and $z=7$.

\subsubsection{Reionization feedback}
\label{sec:models:reionization}
Simulations such as {\sc CoDa II} \citep{ocvirk20} predict a significant flattening or turnover at $M_{\text{UV}} > -15$. This discrepancy is largely driven by efficient photo-heating during reionization, where the rising ionizing background increases gas temperature and suppresses accretion into low-mass halos. Crucially, as highlighted by \citet{dawoodbhoy23}, the impact of radiative feedback is subject to significant environmental variations. Probing different volumes with distinct local reionization histories ranging from early-reionizing voids to late-reionizing overdensities can lead to substantial variations in the predicted number density at the faint end. Our data suggest that the global suppression may be less efficient than these models assume or that our specific survey volumes represent environments where the turnover is shifted to even fainter magnitudes.

The {\sc DRAGONS} (Dark-ages, Reionization, And Galaxy-formation Observables from Numerical Simulations) project is designed to bridge the gap between small-scale galaxy physics and large-scale cosmic structures by using a semi-analytical model built on top of a massive N-body Dark Matter simulation. Among the ingredients, SN feedback is modeled with a strong mass-loading dependency. This prescription results in a significantly more efficient gas ejection in the shallow potential wells of lower-mass halos. In addition, the overall normalization of the LF is highly sensitive to the specific parameterization of the stellar to halo mass relation (SHMR). The simulations predict that once the Universe reaches a certain level of ionization, the heat from the UV background suppresses star formation in halos smaller than $\sim 10^9$\msol. This leads to a predicted turnover in the luminosity function, similar to {\sc CoDa}, driven by the global growth of the UV background.

\subsubsection{Numerical Resolution} Some simulations shown in Figure \ref{fig:lf_models}, most notably {\sc ENZO} \citep{oshea15} and {\sc FIREbox} \citep{feldmann23}, show "wiggles" or a sudden flattening at the extreme faint end. These are likely caused partially or mostly by numerical artifacts rather than physical turnovers. When a simulation reaches its mass resolution limit, it can no longer resolve the internal structure of the smallest halos, leading to an artificial depletion of the galaxy population that does not reflect the physical reality of the $z \sim 7$ Universe.

The {\sc THESAN} project \citep{kannan22} is a  large-volume (95 cMpc on a side) radiation-hydrodynamics (RHD) simulation. The flagship simulation includes sophisticated models for supernova feedback, stellar winds, and AGN feedback. It uses the TNG feedback recipes, which were calibrated to match UVLF at lower redshifts. Therefore, these observations provide a test for the redshift evolution of such models. 
With a star particle mass of $5.8\times 10^{5}$\msol, {\sc THESAN} cannot smoothly form stars in small galaxies. At \muv $= -13$, the total stellar mass is around $10^6$ \msol. This means such a galaxy is composed of only $\sim 10$ star particles. Below this limit, the simulation literally lacks the building blocks to create a galaxy, creating an artificial drop-off in the LF.

{Generally, SN feedback energy is coupled to the mass of the star particles, although this is not the case in some simulations such as {\sc THESAN}}. When a $3 \times 10^5 M_{\odot}$ particle explodes, it releases a massive, discrete burst of energy. In a tiny halo, this single burst is often enough to blow all the gas out of the galaxy instantly. This makes star formation inefficient at the low-mass end. In contrast, with the 10 pc resolution of {\sc Sphinx} ($10^3 M_{\odot}$ particles), that same energy would be released more gradually by smaller clusters, allowing the galaxy to retain some gas.

Ultimately, our results indicate that the turnover magnitude $M_T$ likely occurs at magnitudes fainter than $M_{\text{UV}} = -12.3$. This poses a significant challenge to models that rely on early or aggressive suppression mechanisms to regulate the dwarf galaxy population. Another aspect is that most galaxy formation models often start with a cosmological baryon fraction in low-mass halos, necessitating the implementation of aggressive feedback to reconcile simulated stellar masses with observations. However, if accretion is suppressed by reionization heating, only light feedback is needed to regulate the limited fuel reservoir and suppress star formation to observed levels \citep{mcquinn22}.

\begin{table}
    \centering
    \caption{The binned UVLF data points, including the number of sources in each magnitude bin}
    \begin{tabular}{lcc}
	\hline
    \hline
	M$_{\rm UV}$ & N$_{\rm obj}$  & $\log (\phi)$ (Schechter) \\
       AB      &           &[mag$^{-1}$Mpc$^{-3}$] \\
    \hline 
    $-22.19$ & --  & $-6.00^{+0.48}_{-0.48}$\\
    $-21.69$ & --  & $-4.39^{+0.12}_{-0.14}$\\
    $-21.19$ & --  & $-4.33^{+0.14}_{-0.17}$\\
    $-20.69$ & --  & $-3.70^{+0.08}_{-0.09}$\\
    $-20.19$ & --  & $-3.55^{+0.10}_{-0.12}$\\
    $-19.69$ & --  & $-3.23^{+0.09}_{-0.11}$\\
    $-19.19$ & --  & $-2.93^{+0.13}_{-0.16}$\\
    $-18.69$ & --  & $-2.84^{+0.13}_{-0.16}$\\
    $-18.50$ & 26  & $-2.84\pm 0.20$\\
    $-17.94$ & --  & $-2.08^{+0.11}_{-0.12}$\\
    $-17.50$ & 80  &  $-2.24\pm 0.11$\\
    $-16.94$ & -- & $-2.20^{+0.15}_{-0.19}$\\
    $-16.50$ & 177  & $-1.62\pm 0.12$\\
    $-15.50$ & 51  & $-1.63\pm 0.29$\\
    $-14.50$ & 32  & $-0.87\pm 0.27$\\
    $-13.50$ & 13  & $-0.35\pm 0.34$\\
    $-12.50$ & 6  & $0.03\pm 0.67$\\
	   \hline
        \end{tabular}
\label{tab:binned_lf}
\end{table}

\subsection{Comparison with Semi-Analytical Models}

While hydrodynamical simulations provide a direct look at the baryonic processes of galaxy formation, semi-analytical models (SAMs) allow for the exploration of a broader parameter space regarding feedback and halo growth. As shown on the right panel of Figure~\ref{fig:lf_models}, our observed LF reveals a significant tension with several popular SAM prescriptions, particularly at the faint-end limit of $M_{\text{UV}} \approx -12$.

\subsubsection{Star formation thresholds} A primary feature in many SAMs is the implementation of a strict mass or velocity threshold for star formation. The \citet[][]{yue16} model, utilizing a circular velocity threshold of $v_c^* = 50$ km s$^{-1}$, predicts a sharp cliff in the LF at \muv $\approx -14$. Similarly, the \citet[][]{dayal15} Jeans Mass Feedback model shows a dramatic turnover. These results are in direct conflict with our data, which exhibits a continuous rise, suggesting that star formation efficiency in low-mass halos remains significant even in potential wells significantly shallower than these models assume.
    
Similarly, the {\sc ASTRAEUS} Strong Heating model \citep{hutter21} predicts a much lower normalization and an earlier turnover than our observations. This highlights that SAMs assuming global radiative heating tend to over-quench the dwarf galaxy population. Our results favor models where the ionizing background either acts more slowly or is locally modulated, preventing the immediate suppression of small halos.

\subsubsection{Initial Mass Function Variations} More recent models, such as \citet{Trinca24} and {\sc DELPHI} \citep{mauerhofer25}, investigate the effects of an evolving IMF (eIMF) on the UVLF. However, in both sets of models, the variation between a fiducial and an evolving IMF appears to have a relatively minor impact on the resulting number densities. Notably, the \citet{Trinca24} models, both the fiducial and eIMF versions, already exhibit a higher normalization than our observations across much of the faint end. Perhaps it is not surprising that the \citet{Trinca24} model reaches the highest number densities among the theoretical models at $M_{\text{UV}} \approx -14$ since it was calibrated to reproduce the star formation rate density evolution at lower redshift ($z < 6$) and the quasar luminosity function at $4 < z < 7$. Nevertheless, its existing over-prediction relative to our data suggests that while a top-heavy IMF is a potential mechanism for increasing luminosity, it may not be the primary driver required to reconcile the underlying dark matter halo distribution with the observed low-luminosity galaxy population.

\subsubsection{Time and Mass Resolutions}

The {\sc SHARKS} model (Lagos et al. in prep) is a semi-analytical model implemented on the {\sc SURFS} suite of N-body simulations, specifically designed to investigate the evolution of galaxies and AGNs by incorporating sophisticated treatments of gas cooling, star formation, and multi-wavelength emission. To assess the impact of numerical and temporal resolution on our findings, we compare our results with the {\sc SHARKS} predictions, where the resolution of the underlying dark matter halo merger trees, combined with the frequency of temporal snapshots improves convergence at high redshift. The {\sc SHARKS} intermediate-resolution runs with higher snapshot frequency (270 snapshots) show that fine temporal and mass resolutions are required to better capture the assembly history of low-mass halos. While these configurations successfully reproduce the observed SFRD at $z > 6$ (Lagos et al. in prep), it remains below the observed number densities of low-mass galaxies.

\subsubsection{Empirical Extrapolations} The semi-empirical model of \citet{mason15} provides a framework for connecting galaxy luminosity to dark matter halo assembly. The model is based on a calibration of the SHMR at these redshifts. The model introduces a star formation efficiency that varies with halo mass. This efficiency is calibrated at $z \approx 6$ to match the observed UV luminosity function, which effectively bypasses the uncertainties of sub-grid hydrodynamics. It predicts a persistently steep, slightly shallower than our observations, faint-end slope ($\alpha \sim -1.95$) down to $M_{\text{UV}} \approx -13$, suggesting that the growth of the UV luminosity function is primarily driven by the underlying assembly of dark matter halos. Because the model is empirical and the halos continue to exist down to very low masses, it does not predict a turnover.

The persistence of a steep slope down to $M_{\text{UV}} = -12$ poses a challenge to the standard star formation suppression mechanisms found in SAMs. It necessitates a re-evaluation of how gas is retained and cooled in halos with masses near the atomic cooling limit during the Epoch of Reionization.

\subsection{Comparison with \jwst-inspired models}

The unexpected abundance of UV-bright galaxies at $z > 10$ has prompted a surge in theoretical frameworks aimed at reconciling the bright end of the early-Universe luminosity function with standard cosmological models. Several physical mechanisms have been proposed to explain this excess. The enhanced star formation efficiency model assumes feedback-free starbursts \citep{li24} where the gas is converted into stars with high efficiency ($\epsilon \approx 0.2$--$1.0$) before supernova feedback can regulate the process. The reduced dust attenuation model \citep{ferrara24} assumes that outflows in early systems are able to expel the gas and the dust from the star-forming regions, which allows the UV light to escape unattenuated. The models based on stochastic star formation histories \citep{mirocha23,munoz23,shen23,sun24} use the resulting luminosity fluctuations to boost the number of galaxies at the bright end of the distribution. For these mechanisms to be physically viable, they must not only explain the $z > 10$ excess but also evolve to match the observed galaxy population at lower redshifts and fainter magnitudes.

\begin{figure}
    \centering
    \includegraphics[width=\linewidth]{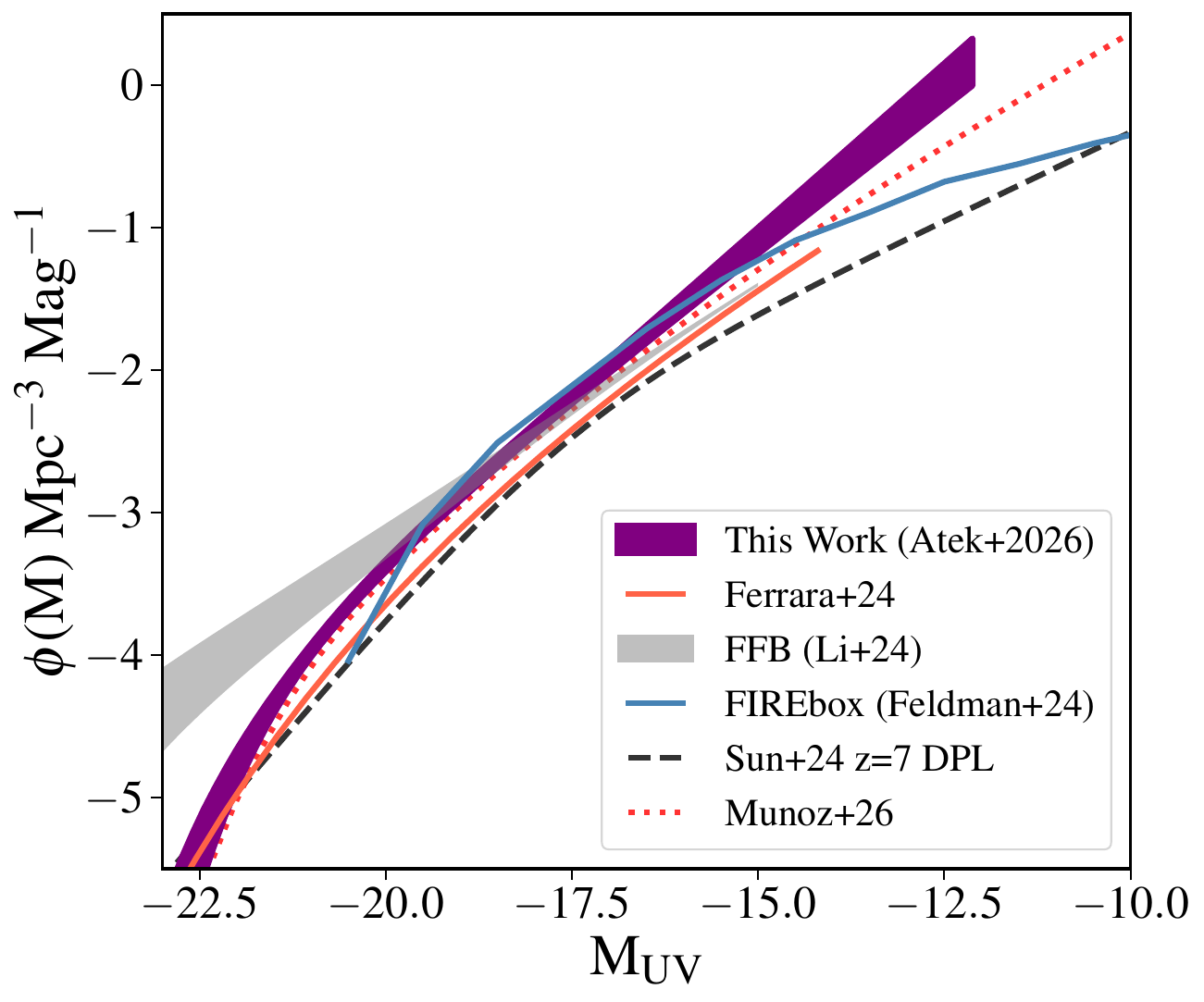}
    \caption{Same as Figure \ref{fig:lf_models} but the comparison is made with the post-\jwst\ models that have been designed to match the most recent observations of the bright end of the UVLF at $z>10$. The feedback-free model \citep{li24}, is represented by the gray curve, the attenuation-free model \citep{ferrara24} by the orange curve, and the {\sc FIREbox} simulations \citep{feldmann23} are shown with the blue solid line. The stochastic SFH models by \citet{sun24} and \citet{munoz26_burst} are shown with a dashed-black curve and a dotted-red curve, respectively. 
    }
    \label{fig:lf_post}
\end{figure}

In Figure \ref{fig:lf_post}, we compare these post-JWST models at $z=7$ against our observed UVLF. The feedback-free model exhibits a significant discrepancy with the data; the high star formation efficiencies required at $z > 10$ lead to a substantial over-prediction of the bright-end density at $z=7$, while simultaneously producing a shallower faint-end slope that extends to $M_{\text{UV}} \approx -15$. Similarly, the attenuation-free model shows a marked normalization offset in both the characteristic magnitude ($M^*$) and density ($\phi^*$). The bursty FIRE star formation scenario \citep{sun2023} fails to reproduce the overall LF shape, significantly under-predicting galaxy number densities across nearly the entire luminosity range at $z=7$. The {\sc FIREbox} simulations do a better job, although the predicted flattening below \muv$=-15$ is in disagreement with our observations.

These persistent disagreements emphasize that while these mechanisms may explain the $z > 10$ bright-galaxy problem, they lack the necessary redshift and mass-dependent evolution to remain consistent with the established galaxy population at the end of reionization. 
An avenue to reconcile these results may be a mass (or redshift) dependence of the physical parameters. For instance, in \citet{munoz26_burst} it was shown that the H$\alpha$/UV ratios observed in $z\sim 6$ \jwst\ galaxies imply a halo-mass-dependent burstiness, which would become stronger at earlier times when halos are less massive. This mass-dependent bursty model agrees better with our measurements (assuming no halo-mass turnover so the UVLF remains steep at the faint end).
This highlights the need for a self-consistent framework that transitions naturally from the early starburst regime to the regulated growth seen in lower-mass galaxies and at later epochs.

\begin{table*}
    \centering
        \caption{The best-fitting parameters and derived densities integrated to $M_{lim} = -12$.}
    \begin{tabular}{lcccccc}
    \hline
    \hline
          & $\log (\phi^*)$  & $M^{*}$  & $\alpha$  & M$_{\rm turn}$& $\beta$ & $\log (\rho_{\rm UV})$  \\
          Reference & [mag$^{-1}$Mpc$^{-3}$]  & [mag]  & &  [mag]& & [ergs s$^{-1}$ Hz$^{-1}$ Mpc$^{-3}$ ] \\
    \hline
    \hline
This Work\footnote{~adopting a Schechter form}   & $-3.68 _{-0.17}^{+0.17}$ & $-21.29_{-0.24}^{+0.22}$ & $-1.98_{-0.05}^{+0.06}$ & -- & -- & $26.34_{-0.09}^{+0.09}$  \\  
This Work\footnote{~a modified Schechter form with a turnover}   & $-3.88 _{-0.23}^{+0.20}$ & $-21.51_{-0.31}^{+0.29}$ & $-2.09 _{-0.06}^{+0.07}$ & $>-12.3$ & $1.26 _{-0.85}^{+1.14}$& $26.21_{-0.12}^{+0.07}$ \\
\citet{atek18}$^{\rm a}$  & $-3.43 _{-0.21}^{+0.21}$ & $-20.74_{-0.20}^{+0.21}$ & $-1.98 _{-0.09}^{+0.11}$ &--&--& $26.34 \pm 0.14$   \\
\citet{atek18}$^{\rm b}$  & $-3.54 _{-0.07}^{+0.06}$ & $-20.84_{-0.30}^{+0.27}$ & $-2.01 _{-0.14}^{+0.12}$ & $-14.93_{-0.52}^{+0.61}$& $0.48 _{-0.25}^{+0.49}$ &  $26.16 \pm 0.13$ \\
\citet{ishigaki18} & $-3.78 _{-0.15}^{+0.15}$ & $-20.89_{-0.13}^{+0.17}$ & $-2.15 _{-0.06}^{+0.08}$ & -- & --&  $26.68 \pm 0.15$ \\
\citet{bouwens22c} & $-3.70\pm0.10$ & $-21.13\pm0.08$ & $-2.05\pm0.06$ &$>-14$ & $0.24\pm0.20$ & $26.24 \pm 0.10$ \\
    \hline
    \end{tabular}
    \label{tab:param}
\end{table*}

\section{the UV luminosity and SFR density}
\label{sec:rhouv}

\begin{figure}
    \centering
    \includegraphics[width=\linewidth]{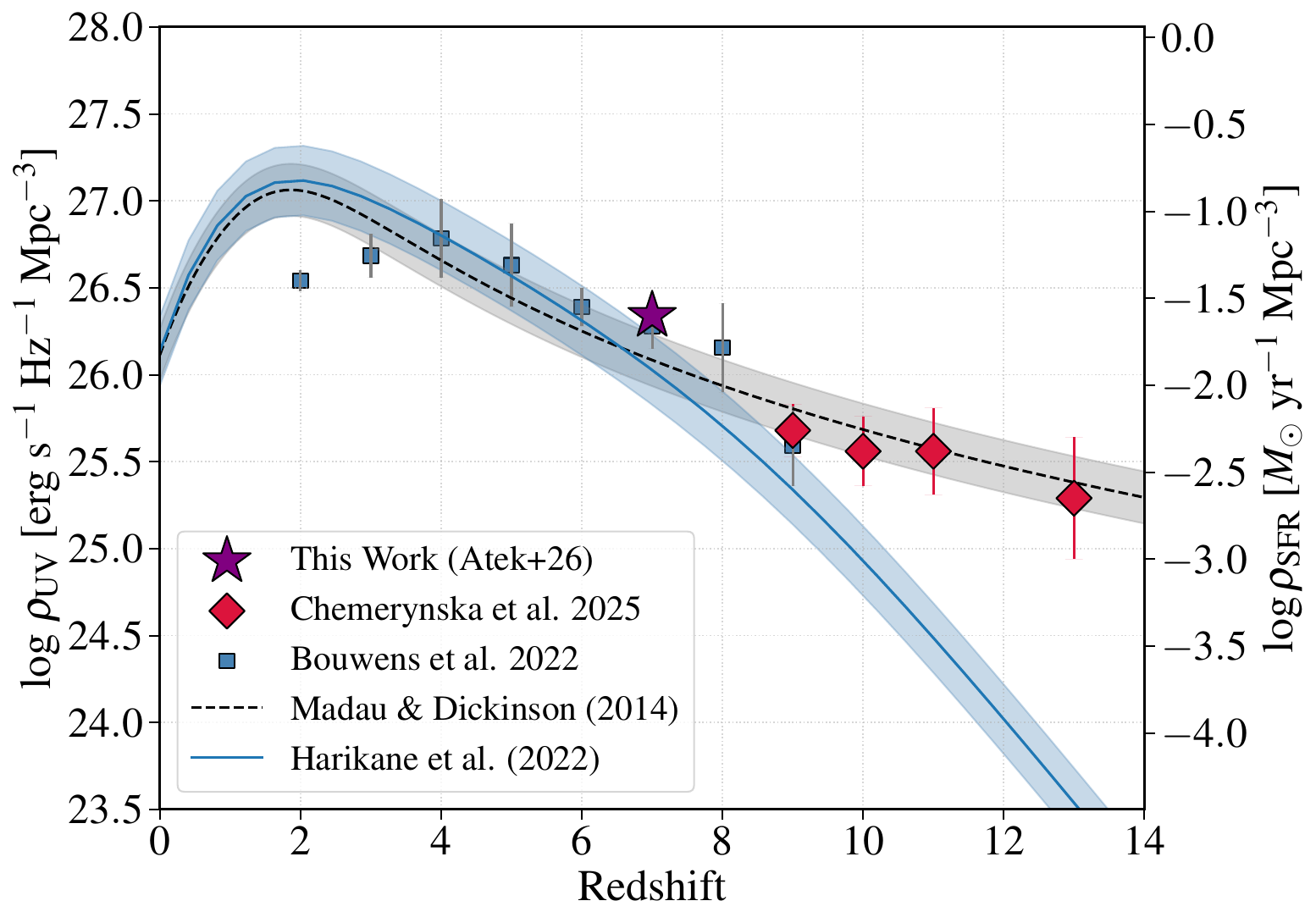}
    \caption{Evolution of the cosmic UV luminosity density ($\rho_{\rm UV}$) and star formation rate density ($\rho_{\rm SFRD}$) as a function of redshift. The results from the \glimpse\ survey at $z>9$ \citep{chemerynska25} are plotted alongside our determinations at $z \approx 7$  and a compilation of literature results. The right-hand vertical axis shows the equivalent $\rho_{\rm SFRD}$ calculated using the conversion factor $\kappa_{\rm FUV} = 1.15 \times 10^{-28}\,M_\odot\,{\rm yr}^{-1} / ({\rm erg}\,{\rm s}^{-1}\,{\rm Hz}^{-1})$. The gray-shaded region represents the functional fit from \citet{madau14}, while the blue-shaded region indicates the empirical evolution determined by \citet{harikane22}. Note that for consistency the \glimpse\ constraints, both present and those of \citet{chemerynska25}, and those of \citet{bouwens22c} are integrated to the same limit of $M_{\rm UV} = -12$ to capture the contribution of the faint-end population. However, the best-fit relations derived by \citet{madau14} and \citet{harikane22} are kept to their original integration limit of $M_{\rm UV} = -17$. 
    }
    \label{fig:rhouv}
\end{figure}

We investigate the evolution of the UV luminosity density ($\rho_{\rm UV}$) by integrating the luminosity function down to a faint-end limit of $M_{\rm lim} = -12$. While previous investigations of the UV luminosity function during the Epoch of Reionization were typically limited to magnitudes brighter than $M_{\rm UV} = -15$, they relied on extrapolations to account for the fainter galaxy population. In this work, we provide the first direct estimate of the faint-end slope reaching these extreme limits, allowing for a more robust calculation of the total UV budget. To contextualize our findings within the broader cosmic star formation history, we convert our $\rho_{\rm UV}$ determinations into equivalent star formation rate (SFR) densities ($\rho_{\rm SFR}$). Following the prescriptions of \citet{madau14} and assuming a \citet{chabrier03} initial mass function (IMF), a constant star formation rate, and a metallicity of $Z = 0.002 Z_{\odot}$, we adopt a conversion factor of $\mathcal{K}_{\rm FUV} = 1.15 \times 10^{-28} M_{\odot} \, {\rm yr}^{-1} \, {\rm erg}^{-1} \, {\rm s} \, {\rm Hz}$. This conversion allows the simultaneous presentation of both SFR and UV luminosity densities in the comparison shown in Figure~\ref{fig:rhouv}. The derived values are also presented in Table \ref{tab:param}. We note that none of these values are corrected for dust attenuation. 

We see that the present \rhouv\ determination at $z\sim 7$ is in agreement with extrapolations of previous determinations that we integrated down to \muv$=-12$ \citep{atek18}. The measurement of \citep{bouwens22} is only slightly lower because their steeper faint-end slope somewhat compensates for the turnover of the LF. Most of the \rhouv\ values are significantly above the redshift evolution determined by \citet[][]{madau14} (dashed gray line), which is based on an integration limit of \muv=$-17$ mag. We note that when integrated to the same magnitude limit, we find Log(\rhouv/ergs~s$^{-1}$ Hz$^{-1}$ Mpc$^{-3}$)=25.95, which is consistent with the determination of \citet{madau14} at the same redshift. It is clear that both the faint-end slope and the truncation limit without a rollover of the LF provide a larger UV budget of galaxies.

\section{implications on cosmic reionization}
\label{sec:reionization}

The cosmic reionization history can be described by the redshift evolution of the ionized volume fraction, $Q_{\text{H II}}$, which is the result of a competition between ionizing photon production and the rate of recombination \citep[e.g.][]{madau99,bolton07,robertson13}:

\begin{equation}
    \dot{Q}_{\text{H II}} = \frac{\dot{n}_{\text{ion}}}{\langle n_{\text{H}} \rangle} - \frac{Q_{\text{H II}}}{t_{\text{rec}}}.
    \label{eq:dq}
\end{equation}

In Equation \ref{eq:dq}, the source term is defined by the production rate of ionizing photons, $\dot{n}_{\text{ion}}$, normalized by the mean number density of hydrogen, $\langle n_{\text{H}} \rangle$. These quantities are derived as follows:

\begin{equation}
    \dot{n}_{\text{ion}} = \int_{-\infty}^{M_{\text{trunc}}} f_{\text{esc}}(M_{\text{UV}}) \xi_{\text{ion}}(M_{\text{UV}}) \Phi(M_{\text{UV}}) L(M_{\text{UV}}) dM_{\text{UV}} \\
    \label{eq:n_ion}
\end{equation}

where the term \fesc\ accounts for the fraction of ionizing photons that escape the galactic environment into the intergalactic medium (IGM), while \xiion\ is the conversion factor between the UV luminosity density and the ionizing photon emission rate, also called the ionizing efficiency. The mean number density is derived as follows:

\begin{equation}
    \langle n_{\text{H}} \rangle = \frac{X_{\text{p}} \Omega_{\text{b}} \rho_{\text{c}}}{m_{\text{H}}},
    \label{eq:n_H}
\end{equation}

where $X_{\text{p}}$ represents the primordial hydrogen mass fraction, $\rho_{\text{c}}$ is the critical density of the Universe, and $m_{\text{H}}$ is the mass of a hydrogen atom.

The sink term in Equation \ref{eq:dq} is governed by the mean gas recombination time, $t_{\text{rec}}$, expressed as:

\begin{equation}
    t_{\text{rec}} = \frac{1}{C_{\text{H II}} \alpha_{\text{B}}(T) (1 + Y_{\text{p}} / 4X_{\text{p}}) \langle n_{\text{H}} \rangle (1 + z)^{3}},
    \label{eq:t_rec}
\end{equation}

where $\alpha_{\text{B}}(T)$ denotes the Case-B recombination coefficient at IGM temperature $T$, and $Y_{\text{p}}$ is the primordial mass fraction of helium. The clumping factor, $C_{\text{H II}} \equiv \langle n_{\text{H II}}^{2} \rangle / \langle n_{\text{H II}} \rangle^{2}$, accounts for the local density fluctuations of ionized hydrogen \citep[e.g.][]{finlator12,davies24}

\subsection{The ionizing emissivity}

\begin{figure}
    \centering
    \includegraphics[width=\linewidth]{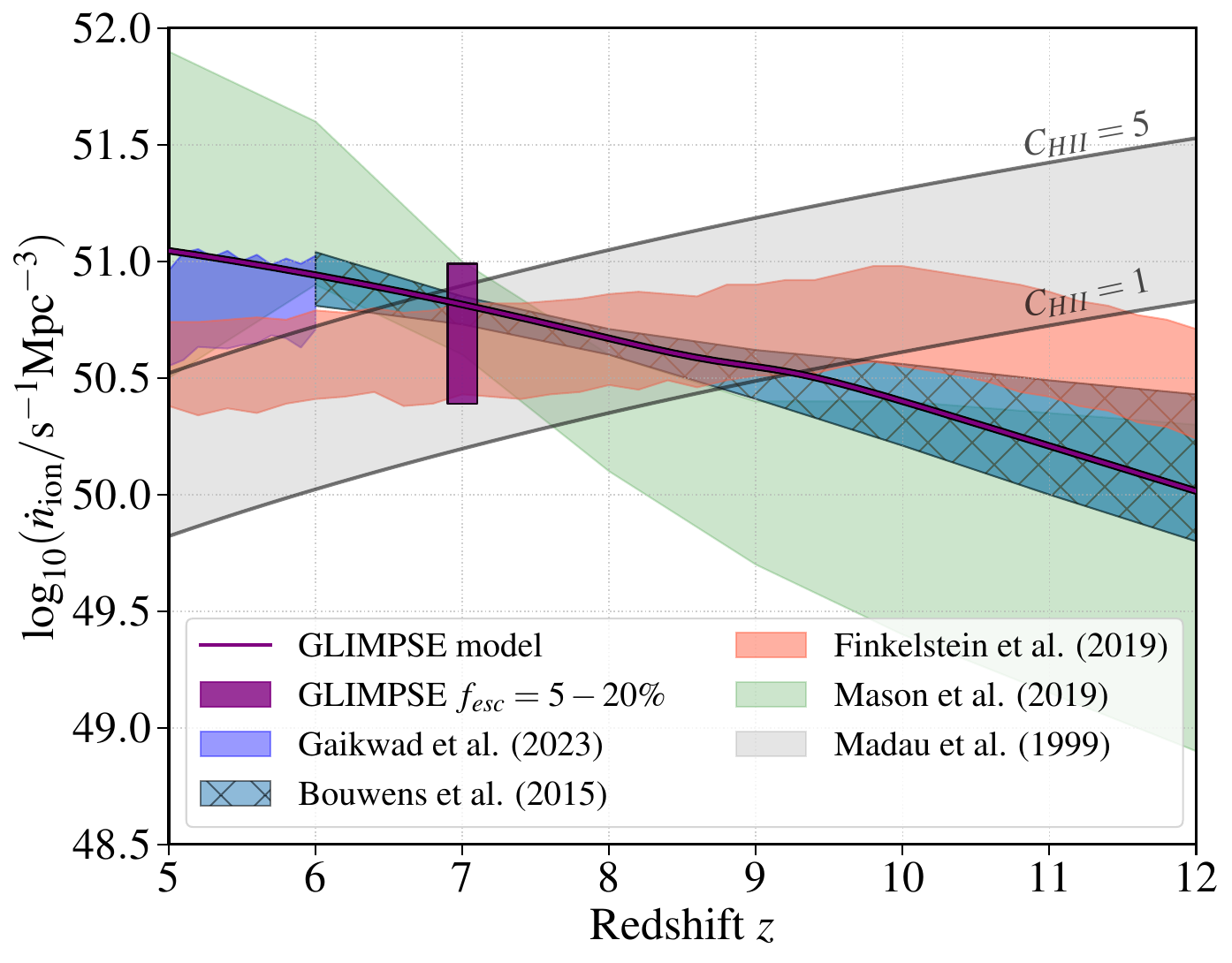}
    \caption{Evolution of the comoving ionizing photon emissivity rate density, $\log_{10}(\dot{n}_{\text{ion}} / \text{s}^{-1} \text{Mpc}^{-3})$, as a function of redshift $z$ in the range $5 < z < 12$. The purple data point represents the results from this work, compared against various observational constraints and theoretical models, including the \glimpse\ model (purple line), \citet{gaikwad23} (blue shaded region), \citet{bouwens15} (hatched blue region), and \citet{finkelstein19} (red shaded region). Theoretical reionization limits from \citet{madau99} and \citet{mason19} are shown in grey and green shaded regions, respectively. The gray shaded region indicates the emissivity required to maintain reionization for typical clumping factors between $C_{\text{HII}} = 1$ and $C_{\text{HII}} = 5$.
    }
    \label{fig:nion}
\end{figure}

Using the derived $\rho_{\text{UV}}$, we compute the ionizing photon rate \nion, by adopting the \glimpse\ constraints on the ionizing properties of these faint galaxies: a production efficiency \xiion\ from Chisholm et al. (in prep) and an escape fraction \fesc\ from \citet{jecmen26}. In Chisholm et al. (in prep), we used medium bands to measure the \ha\ emission in $z\sim6-6.5$ galaxies and infer \xiion\ in galaxies down to \muv=$-16$. The best-fit relation as a function of magnitude at $z\sim 6$ closely follows the result of \citet{simmonds24b}: 

$\log_{10} \left(\xi_{\text{ion}}(z, M_{\text{UV}}) \right) = (0.003 \pm 0.003)z + (-0.018 \pm 0.003)M_{\text{UV}} + (25.984 \pm 0.053).$

At $z=7$, this calculation yields a value of log(\nion / s$^{-1}$ Mpc$^{-3}) = 50.85$. Adopting an escape fraction in the range \fesc=$5-20\%$ results in the purple-shaded vertical stripe at $z=7$ shown in Figure \ref{fig:nion}. Using the redshift-evolution of these relations across the range $5 < z < 12$, we derive the \glimpse\ model (purple line). The inferred \nion\ at $z=7$ is slightly higher than the literature results, such as \citet{bouwens15} (hatched blue) and \citet{finkelstein19} (red shaded region), which typically relied on shallower UVLF integrations or more conservative assumptions on the ionizing properties. The gray lines represent the critical emissivity required to balance recombinations and maintain reionization for different IGM clumping factors ($C_{\text{HII}}$). Our $z=7$ point sits well above the $C_{\text{HII}} = 1$ limit and is consistent with the higher $C_{\text{HII}} = 5$ requirement. As shown in the figure, if the IGM is highly clumped ($C_{\text{HII}} = 5$), the required emissivity is more than 0.5 dex higher than the $C_{\text{HII}} = 1$ case.

Because our \nion\ remains elevated at higher redshifts (the purple curve), it suggests that the faintest galaxies (\muv$> -15$) provide a sufficient ionizing budget to sustain reionization. We note that the \glimpse\ model remains consistent with, or even falls below, certain literature results, particularly at $z > 8$. This divergence arises primarily because previous studies often extrapolated the UVLF to faint magnitudes while adopting a significantly higher fiducial escape fraction of \fesc$= 20\%$.

\subsection{Reionization history}

In the bottom panel of Figure \ref{fig:reionization}, adapted from \citet{munoz24}\footnote{\url{https://github.com/JulianBMunoz/Simple-Reionization-Plot}}, we present the redshift evolution of the ionized hydrogen volume fraction, $Q_{\text{HII}}$, across our different modeling scenarios. For context, we include a suite of empirical constraints derived from IGM opacity studies \citep{nakane23}, including quasar absorption line spectra \citep{mcgreer15, greig17, greig18, davies18, banados18, wang20} and the statistics of \lya\ emitters \citep{sobacchi15, hoag19, mason19}. Initially, we consider a fiducial reionization model. In this canonical case, the escape fraction $f_{\text{esc}}$ and the ionizing efficiency $\xi_{\text{ion}}$ are assumed to be independent of the absolute UV magnitude, $M_{\text{UV}}$. This allows the source term in Equation \ref{eq:n_H} to be governed by the magnitude-averaged product $\langle f_{\text{esc}} \xi_{\text{ion}} \rangle$. We utilize the redshift-evolving UVLF from \citet{bouwens21}, integrated down to a truncation magnitude of \muv$= -13$. Adopting standard parameters of $\log(\xi_{\text{ion}}/\text{erg}^{-1}\text{Hz}) = 25.2$ and $f_{\text{esc}} = 0.2$ \citep[][]{robertson15}, the resulting reionization history (black solid curve) demonstrates broad consistency with the observational data (blue circles), showing a transition to a fully ionized IGM by $z \approx 6$.

Beyond local IGM probes, we characterize our models by plotting the integrated Thomson scattering optical depth of CMB photons, $\tau_{\text{e}}$ in the top panel of Figure \ref{fig:reionization}. This parameter provides a global integral of the ionization history and is defined as:

\begin{equation}
    \tau_{\text{CMB}}(z) = c \sigma_{\text{T}} \int_{0}^{z} \frac{n_{\text{e}}(z') (1+z')^{2}}{H(z')} dz',
\end{equation}

where $c$ is the speed of light, $\sigma_{\text{T}}$ is the Thomson scattering cross-section, and $H(z)$ is the Hubble parameter. The comoving electron number density, $n_{\text{e}}(z)$, accounts for the contributions from both ionized hydrogen and helium, and assuming the second ionization of helium at $z=4$, it corresponds to $ n_{\text{e}}(z) = Q_{\text{HII}}(z) \langle n_{\text{H}} \rangle $. This integrated measurement is particularly sensitive to the high-redshift tail of reionization, which provides an important reference for reionization models. The latest \textit{Planck} constraints indicate $\tau_{\text{CMB}} = 0.054 \pm 0.007$ \citep{planck18}, which represented by the red-shaded region in Figure \ref{fig:reionization}. We can see that the canonical reionization model produces an optical depth in good agreement with the {\em Planck} constraints. 

The most recent \jwst\ observations have updated many of these standard assumptions on the ionizing properties of early galaxies, while the present work is providing unprecedented constraints on their number density. Observations of faint galaxies (\muv$\gtrsim -16$) by \citet{Atek24a} indicate significantly elevated ionizing efficiencies, with values reaching $\log_{10}(\xi_{\text{ion}} / \text{erg}^{-1}\text{Hz}) \approx 25.8$. Such high efficiencies imply that the faint end of the UVLF may have played a more dominant role in reionization than previously modeled. Furthermore, \citet{prietolyon23} and \citet{simmonds24a} have proposed empirical relations between the ionizing efficiency and the absolute UV magnitude. The resulting $\tau_{\rm CMB}$ would likely exceed the \textit{Planck} legacy constraint. This suggests that the escape fraction \fesc\ may be significantly lower than $0.2$, or that it must decrease in the same faint galaxies where $\xi_{\text{ion}}$ is most elevated \citep{munoz24}. Alternatively, the UVLF must be truncated at brighter magnitudes to avoid ionizing the Universe too early. 

\subsection{The \glimpse\ reionization model}

In addition to the present work, which provides the first constraints on the contribution of the faintest galaxies to the total UV budget, the \glimpse\ survey has also explored the ionizing properties of EoR galaxies. In \citet{jecmen26}, we also derived a relation between the UV slope $\beta$ and the UV magnitude down to \muv=$-12.5$ at $z>6$, including the sample used in the present work, which is best described by a double power law of the form:

\begin{equation}
    \beta(M) = \frac{\beta_{0}}{\left[ 10^{-0.4(M-M^{*})\alpha_{1}n} + 10^{-0.4(M-M^{*})\alpha_{2}n} \right]^{1/n}}
\end{equation}

where $\beta_{0} = -2.45 \pm 0.14$, $M^{*} = -18.5 \pm 1.00$ is the transition magnitude between the two power laws, $n = 11.5 \pm 16.6$, and $\alpha_{1} = 0.11 \pm 0.06$ and $\alpha_{2} = -0.01 \pm 0.02$ are the bright- and faint-end slopes, respectively.

\begin{figure}
    \centering
    \vspace{0.3cm}
    \includegraphics[width=0.99\linewidth]{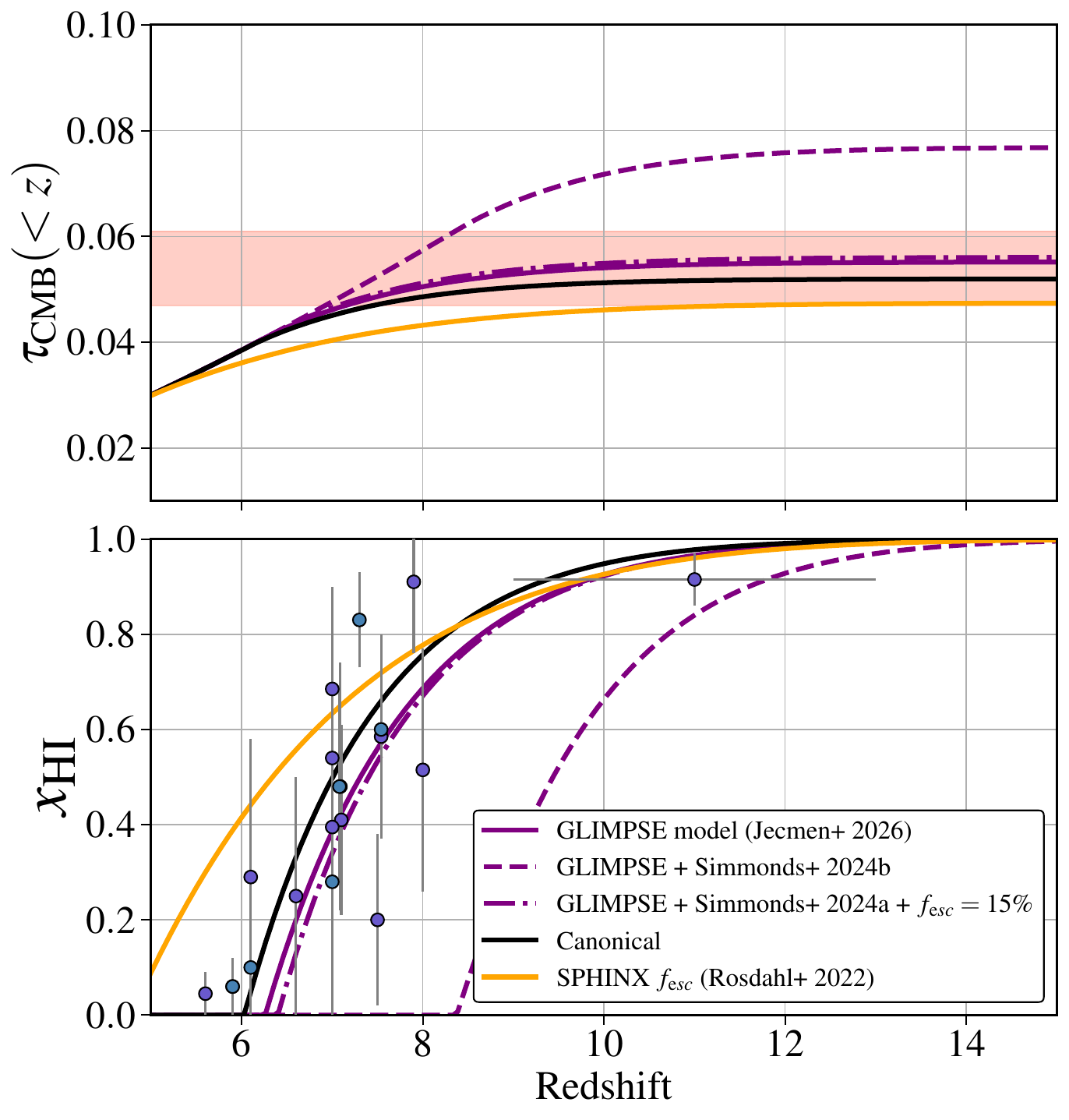}
    \caption{Cosmic reionization histories and CMB optical depth constraints.
    \textbf{Bottom panel:} The evolution of the volume-weighted hydrogen neutral fraction $x_{\text{HI}}$ as a function of redshift. Data points with error bars represent various literature constraints from Ly$\alpha$ emitters (LAEs) and QSO absorption wings \citep{fan06, sobacchi15, greig17,banados18, davies18, mason19, hoag19, wang20}. The purple line shows the prediction from the \glimpse\ model based on updated UVLF \citep[this work and][]{chemerynska25}, \xiion\ (Chisholm et al. in prep), and \fesc\ constraints \citep{jecmen26}. The black line represents the canonical model using standard parameters (log \xiion $= 25.2$, \fesc\ $= 0.2$, and \muv\ limit of $-13$) following \citet{bouwens21} and \citet{robertson15}. The \glimpse\ model suggests a significantly earlier reionization compared to the canonical model. The orange line substitutes \glimpse\ \fesc\ constraints with the {\sc Sphinx}-derived relation of \fesc\ as a function of redshift \citep{rosdahl22}. 
    \textbf{Top panel:} Evolution of the inferred Thomson scattering optical depth $\tau_{\text{CMB}}(<z)$ as a function of redshift. The horizontal pink-shaded region indicates the $1\sigma$ observational constraints from \textit{Planck} observations. 
    }
    \label{fig:reionization}
\end{figure}

In combination with the \fesc-$\beta$ relation from \citet{chisholm22}: $f_{\text{esc}}(\text{LyC}) = (1.3 \pm 0.6) \times 10^{-4} \times 10^{(-1.22 \pm 0.1)\beta}$, this provides a relation between \fesc\ and \muv. Together, these constraints predict a total ionizing emissivity that drives reionization timing in agreement with the IGM opacity measurements. As illustrated by the purple curve in Figure \ref{fig:reionization}, the \glimpse\ model indicates a predominantly ionized IGM ($\bar{x}_{\text{HI}} \ll 1$) by $z \approx 6$.

In addition to the \glimpse\ constraints derived in Chisholm et al. (in prep), multiple recent works have focused on the ionizing efficiency at $z > 6$, which has been derived from both direct spectroscopic measurements \citep[e.g.,][]{Atek24a, llerena25, pahl25,papovich26} and medium-band imaging \citep{prieto23, simmonds24b}. However, beyond photometric uncertainties, the inherently small spectroscopic sample sizes and the limited volumes probed by gravitational lensing surveys raise concerns regarding the representativeness of these estimates for the broader galaxy population. Adopting, for instance, the relation by \citet{simmonds24a} produces an early reionization by $z\sim 8$ (purple dashed curve in Figure \ref{fig:reionization}). If these assumptions are maintained, a reduction in the escape fraction to \fesc$\approx 0.05$ would be required to reconcile the predicted reionization history with the observed CMB optical depth.  However, a higher-magnitude cutoff in the UVLF (e.g., $M_{\text{UV}} < -15$) appears unlikely given our current results and the deep constraints provided by \citet{chemerynska25} $z>9$. 

The most significant source of uncertainty in this model remains the escape fraction of ionizing radiation. Recent efforts have used low-redshift LyC leakers as local analogs to calibrate indirect indicators of \fesc\ for application during the Epoch of Reionization. Results from the Low-z Lyman Continuum Survey \citep[LzLCS][]{flury22a, flury22b}, for instance, suggest a correlation between \fesc\ and the UV continuum slope \citep[$\beta$; ][]{chisholm22}, implying that faint galaxies at $z=7$ possess relatively high escape fractions, averaging approximately $15\%$ \citep{jecmen26}. However, using a different relation between the UV continuum slope and \muv\ derived by \citet{bouwens14} leads to a high ionizing budget, in tension with CMB constraints. To alleviate the tension, significantly lower ionizing efficiency on the order of  $\log_{10}(\xi_{\text{ion}} / \text{erg}^{-1}\text{Hz}) \approx 24.9$ would be required (cf. Table \ref{tab:reionization}). Using the \fesc\ vs redshift relation derived in the {\sc Sphinx} simulations \citep{rosdahl22} lead to a late reionization scenario due to relatively low escape fractions at high$-z$, regardless of the BPASS SED models used in their study. More recently, \citet{papovich26} reported an increase in the ionizing efficiency, albeit with a large scatter, of fainter galaxies at the epoch of reionization. However, they also find relatively small escape fractions around 2-3\%, which significantly reduces the tension in the reionization history.

Overall, these discrepancies highlight a fundamental point in our understanding: either the ionizing photon production and \fesc\ relations follow the most recent determinations from the \glimpse\ data, or the physical mechanisms regulating the IGM, such as the clumping factor ($C_{\text{HII}}$), undergo significant evolution at high redshift. By connecting the latest mean free path measurements from quasar spectra at $z \simeq 6$ to the recombination rate, \citet{davies24b} suggest that the global clumping factor may be as high as $C \approx 12$. This is much higher than previous determinations and implies that galaxies must produce roughly twice as many ionizing photons to complete reionization by $z \sim 6$ (see also~\citealt{cain25}). On the other hand, using \jwst/NIRCam $H_{\text{II}}$ constraints, \citet{austin26} argue that a recombination-weighted clumping factor of $C_{\text{HII, rec}} \approx 6.2$ is required to keep the Universe reionized at $z \simeq 6$ given the prolific photon production seen by \jwst. These higher clumping factors provide a critical sink for ionizing radiation, potentially accommodating the high \nion\ values based on some empirical relations \citep[e.g.][]{mascia24}. For instance, adopting the \citet{simmonds24a} relation, one needs a clumping factor nearly as high as $C_{\text{HII, rec}} \approx 15$ to reconcile the reionization history with $x_{\text{HI}}$ constraints. As discussed in Section \ref{sec:models:reionization}, the impact of local reionization on the number density of faint galaxies is intrinsically tied to the spatial distribution of the IGM. This leads to an environmental dependence in the timing of reionization. Low-mass galaxies residing in dense, highly clumped environments are likely to experience early photo-heating, which truncates their gas accretion and star formation \citep[e.g.][]{mcquinn24}. Consequently, this differential suppression suggests that the faint end of the UVLF may be affected by these environmental effects.   

Alternatively, earlier reionization with a lower CMB optical depth can alleviate recent cosmological hints of dynamical dark energy found in large-scale structure surveys. By omitting low-$\ell$ polarization data from their analysis, \citet{sailer26} infer a significantly higher CMB optical depth of $\tau\approx 0.09$, in contrast to the standard value of $\tau\approx 0.06$ reported by the {\em Planck} collaboration. This elevated value of $\tau$ has notable cosmological implications. Specifically, it helps alleviate some of the recent tensions observed between CMB measurements and Baryon Acoustic Oscillation (BAO) data from the Dark Energy Spectroscopic Instrument \citep[DESI;][]{desi25}.

\begin{table}[h]
\centering
\caption{Comparison of reionization model parameters and the adjustments required for CMB/IGM consistency.}
\label{tab:reionization}
\begin{tabular}{cccc}
\hline \hline
Parameter & \glimpse\ & Canonical & Reconciled Value\footnote{Individual parameter value required to match Planck $\tau_{\rm CMB}$ and IGM opacity constraints while adopting higher \xiion\ and/or \fesc\ values (see text for details).} \\ \hline
$M_{\text{lim}}$              & $-12$           & $-13$     & $\sim -15$           \\
$\log_{10}(\xi_{\text{ion}})$ & $25.2-25.6$     & $25.2$    & $\sim 24.9$     \\
$f_{\text{esc}}$              & $0.15-0.20$     & $0.20$    & $\sim 0.05$     \\
$C_{\text{HII}}$              & $3$             & $3$       & $ 15$               \\ \hline
\end{tabular}
\end{table}

\section{Summary}
\label{sec:summary}

In this work, we have used the intrinsically deepest observations of the distant Universe obtained by the \glimpse\ survey to constrain the prevalence of the faintest galaxies at $z=6-9$. In addition to deep \hst\ imaging from the Frontier Fields program, the \glimpse\ survey has obtained ultra-deep NIRCam imaging of the AS1063 galaxy cluster in 7 broadbands and two medium bands, down to a limiting magnitude of $\sim 30.9$ \citep{atek25}. By leveraging gravitational lensing and high-sensitivity \jwst\ observations, we have characterized the galaxy population down to an absolute magnitude of \muv $\approx -12$. Our primary findings and their implications for both galaxy formation models and cosmic reionization are summarized as follows.

\begin{itemize}

    \item Our UVLF determination incorporates a comprehensive analysis of uncertainties, including statistical errors in magnification, photometry, and redshifts propagated via Monte Carlo perturbations. We also account for lensing systematics by using three different lensing models, which dominate the error budget at the faint end. We find that the $z \sim 7$ UVLF continues to rise steeply down to $M_{\text{UV}} = -12$, roughly three magnitudes deeper than previous robust constraints. The best-fit faint-end slope, $\alpha = -1.98 _{-0.05}^{+0.06}$, shows no evidence of a turnover in the regime where previous \hst\ studies suggested a potential flattening ($M_{\text{UV}} \approx -14$) \citep{atek18, bouwens22c}. When allowing for a turnover parameter ($\beta$) within uncertainties, our results indicate that any curvature is significantly shallower than previously reported, with the turnover magnitude $M_{\mathrm T}$ likely pushed to magnitudes fainter than $-12.3$.

    \item Our observed $z \sim 7$ UV luminosity function and the steep faint-end slope provide a stringent test for theoretical models. We find that hydrodynamical simulations emphasizing bursty stellar feedback (e.g., {\sc Sphinx}, {\sc FIREbox}) better match the observed slope than those implementing aggressive global radiative feedback (including reionization background) (e.g., {\sc CoDa II}), which predict a premature flattening or turnover. Similarly, we find significant tension with semi-analytical models (SAMs) that utilize strict mass or velocity thresholds for star formation \citep[e.g.][]{yue16,dayal15}, as these models under-predict the abundance of faint sources. The lack of an observed turnover suggests that star formation in low-mass halos is more resilient to the ionizing background and/or feedback mechanisms are less efficient than previously assumed.

    \item By combining our deep UV luminosity function with the latest constraints on ionizing production efficiency \xiion\ and escape fractions \fesc\ from \citet{jecmen26}, we derive a comoving ionizing emissivity rate density of log(\nion / s$^{-1}$ Mpc$^{-3}$) $\approx 50.85$ at $z=7$. This value is significantly higher than previous canonical estimates, driven by the prolific contribution of the faint galaxy population. 
    
    \item Our fiducial \glimpse\ model predicts a Universe that is predominantly reionized by $z \sim 6$, although a few recent studies have hinted to a later reionization \citep{bosman22,zhu22}. While this demonstrates that faint galaxies provide a sufficient photon budget to drive reionization, there remain important uncertainties on these empirical relations. Given the high \xiion\ measurements in some spectroscopic and photometric surveys, adopting the \xiion-\muv\ relation derived in \citet{simmonds24a} leads to an accelerated timeline of reionization, which stands in tension with global observational constraints, including the Planck Thomson scattering optical depth ($\tau_{\text{CMB}}$) and IGM neutral fraction measurements if the escape fraction is reduced to \fesc $\lesssim 5\%$. Similarly, adopting higher measurements of \fesc\ we find that the model can be reconciled with global constraints if the ionizing efficiency is lowered to $\log_{10}(\xi_{\text{ion}} / \text{erg}^{-1}\text{Hz}) \approx 24.9$. Since our results now firmly rule out a truncation of the LF at bright magnitudes \muv $\sim -15$, the tension with reionization constraints must reside in the ionizing physics. Specifically, the \xiion\ and \fesc\ values adopted here are still subject to uncertainties, likely due to selection effects in spectroscopic targets and the ongoing challenges of calibrating indirect indicators of ionizing photon escape. Alternatively, the photon budget crisis can be resolved through sink-side physics. Drawing on recent results, an increased IGM clumping factor ($C_{\text{HII}} \sim 6\text{--}12$) could provide the necessary recombination sink to balance the high emissivity observed in some studies. Such elevated clumping would allow for high ionizing production while maintaining a late-reionization history consistent with the IGM opacity.
\end{itemize}

Our results highlight the importance of probing the faint galaxy population to constrain the baryonic physics implemented in the galaxy formation models. Ultimately, the prevalence of faint galaxies imposes limits on the nature and the strength of the radiative feedback. It also forces us to reflect on the differences that exist between observations and cosmological simulations in identifying faint galaxies and constraining their UV luminosity and how this relates to their underlying stellar mass distribution. This is out of the scope of the present paper but will be discussed in a dedicated study. Our results also highlight fundamental uncertainties in the properties of high-redshift galaxies and their impact on the global state of the IGM. Progress on this front will require further high-resolution hydrodynamic simulations of IGM clumping and larger, unbiased spectroscopic surveys to confirm if the ionizing properties of the faint population are truly representative of the cosmic average.

\section*{Acknowledgments}

We thank Raffaella Schneider and Claudia Lagos for stimulating discussions and for sharing their models. HA and IC acknowledge support from CNES, focused on the JWST mission, and the Programme National Cosmology and Galaxies (PNCG) of CNRS/INSU with INP and IN2P3, co-funded by CEA and CNES. IC acknowledges funding support from the Initiative Physique des Infinis (IPI), a research training program of the Idex SUPER at Sorbonne Universit\'e. HA acknowledges support by the French National Research Agency (ANR) under grant ANR-21-CE31-0838. This work has made use of the \texttt{CANDIDE} Cluster at the \textit{Institut d'Astrophysique de Paris} (IAP), made possible by grants from the PNCG and the region of Île de France through the program DIM-ACAV+, and the Cosmic Dawn Center and maintained by S. Rouberol. 

This work is based on observations obtained with the NASA/ESA/CSA \textit{JWST} and the NASA/ESA \textit{Hubble Space Telescope} (HST), retrieved from the \texttt{Mikulski Archive for Space Telescopes} (\texttt{MAST}) at the \textit{Space Telescope Science Institute} (STScI). STScI is operated by the Association of Universities for Research in Astronomy, Inc. under NASA contract NAS 5-26555. The data described here may be obtained from
\url{https://doi.org/10.17909/4byn-fe55}.

Support for US investigators in program \#3293 was provided by NASA through a grant from the Space Telescope Science Institute, which is operated by the Association of Universities for Research in Astronomy, Inc., under NASA contract NAS 5-03127.

\bibliography{references,reference2}{}
\bibliographystyle{aasjournal}

\appendix

\begin{figure}
    \centering
    \vspace{0.4cm}
    \includegraphics[width=\linewidth]{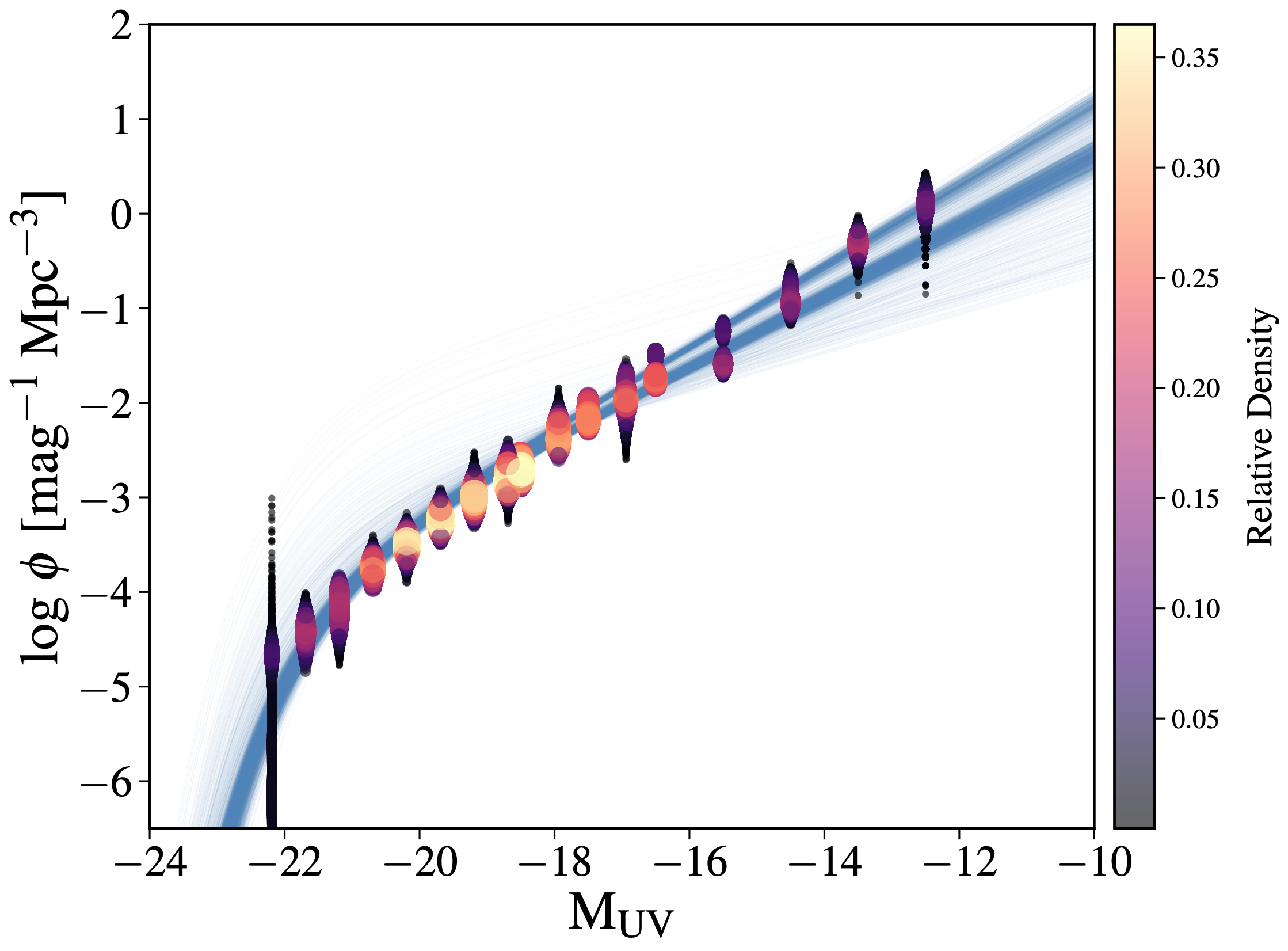}
    \caption{The rest-frame UV luminosity function at $z \sim 7$ derived from the \glimpse\ survey. The colored data points represent the individual realizations from the MCMC sampling, color-coded by their posterior density to illustrate the range of possible UVLF solutions. This distribution accounts for both statistical uncertainties and systematic errors arising from the choice of various lensing models. The shaded blue region and lines represent the associated realizations of the best-fit UVLF.}
    \label{fig:lf_mcmc}
\end{figure}

To provide a more comprehensive view of the uncertainties inherent in our analysis, we include a detailed visualization of our parameter estimation process in this appendix. Figure \ref{fig:lf_mcmc} illustrates the UVLF at $z \sim 7$, specifically highlighting the distribution of individual realizations from our MCMC sampling. By color-coding these realizations according to their posterior density, the figure shows the interplay between statistical noise and the systematic variations introduced by different lensing models. This emphasizes the robustness of our best-fit UVLF, represented by the shaded regions, and provides a better look at the range of possible fits that remain consistent with our observational data.

To complement the visual representation of the MCMC realizations, we also provide a direct comparison of our results based on different lensing models. Table \ref{tab:binned_lf_models} presents the binned UV luminosity functions as derived under three lensing models. This breakdown highlights one of the main systematic uncertainties affecting our UVLF results.

\begin{table*}
\centering
\caption{Binned UV luminosity functions derived for three different lensing models}
\label{tab:binned_lf_models}
\begin{tabular}{c c c c}
\hline
 \muv & \multicolumn{3}{c}{$\log(\phi)$} \\
  mag & \multicolumn{3}{c} {Mpc$^{-3}$ mag$^{-1}$}\\
\\ 
 & \citet{beauchesne2024} & Furtak et al. in prep & \citet{richard14} \\
 
\hline
$-22.19$ & $-6.00 \pm 0.87$ & $-6.00 \pm 0.87$ & $-6.00 \pm 0.87$ \\
$-21.69$ & $-4.39 \pm 0.12$ & $-4.39 \pm 0.12$ & $-4.39 \pm 0.12$ \\
$-21.19$ & $-4.33 \pm 0.14$ & $-4.33 \pm 0.14$ & $-4.33 \pm 0.14$ \\
$-20.69$ & $-3.70 \pm 0.08$ & $-3.70 \pm 0.08$ & $-3.70 \pm 0.08$ \\
$-20.19$ & $-3.55 \pm 0.10$ & $-3.55 \pm 0.10$ & $-3.55 \pm 0.10$ \\
$-19.69$ & $-3.23 \pm 0.09$ & $-3.23 \pm 0.09$ & $-3.23 \pm 0.09$ \\
$-19.19$ & $-2.93 \pm 0.12$ & $-2.93 \pm 0.12$ & $-2.93 \pm 0.12$ \\
$-18.69$ & $-2.84 \pm 0.13$ & $-2.84 \pm 0.13$ & $-2.84 \pm 0.13$ \\
$-17.94$ & $-2.24 \pm 0.11$ & $-2.24 \pm 0.11$ & $-2.24 \pm 0.11$ \\
$-16.94$ & $-2.08 \pm 0.15$ & $-2.08 \pm 0.15$ & $-2.08 \pm 0.15$ \\
\hline
$-18.50$ & $-2.87 \pm 0.16$ & $-2.70 \pm 0.16$ & $-2.84 \pm 0.16$ \\
$-17.50$ & $-2.25 \pm 0.10$ & $-1.99 \pm 0.10$ & $-2.21 \pm 0.10$ \\
$-16.50$ & $-1.68 \pm 0.08$ & $-1.50 \pm 0.08$ & $-1.63 \pm 0.08$ \\
$-15.50$ & $-1.67 \pm 0.10$ & $-1.29 \pm 0.11$ & $-1.62 \pm 0.10$ \\
$-14.50$ & $-0.93 \pm 0.21$ & $-0.69 \pm 0.18$ & $-0.90 \pm 0.21$ \\
$-13.50$ & $-0.35 \pm 0.37$ & $-0.33 \pm 0.29$ & $-0.37 \pm 0.35$ \\
$-12.50$ & $+0.02 \pm 0.67$  & $+0.05 \pm 0.64$  & $+0.03 \pm 0.68$  \\
\hline
\end{tabular}
\end{table*}

\end{document}